\documentclass{article}
\usepackage{graphicx}

\pdfoutput=1

\usepackage{jheppub}
\usepackage{amsmath}
\usepackage{epsfig,multicol,bbm}
\usepackage{url}

\usepackage{graphicx}
\usepackage{setspace}
\usepackage{amssymb}
\usepackage{listings}
\usepackage{epstopdf}
\usepackage{float}
\usepackage{color}
\usepackage{amsmath}
\usepackage{ dsfont }
\usepackage{slashed}
\usepackage{subfigure}

\newcommand{\newc}{\newcommand}
\newc{\gsim}{\lower.7ex\hbox{$\;\stackrel{\textstyle>}{\sim}\;$}}
\newc{\lsim}{\lower.7ex\hbox{$\;\stackrel{\textstyle<}{\sim}\;$}}
\newc{\gev}{\,{\rm GeV}}
\newc{\mev}{\,{\rm MeV}}
\newc{\ev}{\,{\rm eV}}
\newc{\kev}{\,{\rm keV}}
\newc{\tev}{\,{\rm TeV}}

\newc{\mz}{M_Z}
\newc{\mpl}{M_*}
\newc{\mw}{m_{\rm weak}}
\newc{\nr}[1]{N^c_R{}_{#1}}
\usepackage{amsmath}

\def\beq{\begin{equation}}
\def\eeq{\end{equation}}
\def\bea{\begin{eqnarray}}
\def\eea{\end{eqnarray}}
\def\bitem{\begin{itemize}}
\def\eitem{\end{itemize}}
\newc{\ie}{{\it i.e.}}          \newc{\etal}{{\it et al.}}
\newc{\eg}{{\it e.g.}}          \newc{\etc}{{\it etc.}}
\newc{\cf}{{\it c.f.}}

\newcommand\fverb{\setbox\fverbbox=\hbox\bgroup\verb}
\newcommand\fverbdo{\egroup\medskip\noindent%
            \fbox{\unhbox\fverbbox}\ }
\newcommand\fverbit{\egroup\item[\fbox{\unhbox\fverbbox}]}
\newbox\fverbbox

\title{A Complete Model of Low-Scale Gauge Mediation}

\date{\today}

\author[a,b]{Nathaniel Craig,}
\author[a]{Simon Knapen,}
\author[a]{David Shih,}
\author[a]{Yue Zhao}

\affiliation[a]{Department of Physics, Rutgers University \\
Piscataway, NJ 08854 }
\affiliation[b]{ School of Natural Sciences, Institute for Advanced Study \\
Princeton, NJ 08540}

\preprint{RUNHETC-2012-13}

\abstract{Recent signs of a Standard Model-like Higgs at 125 GeV point towards
large $A$-terms in the MSSM. This presents special challenges for gauge mediation, which by itself predicts vanishing $A$-terms at the messenger scale. In this paper, we review the general problems that arise when extending gauge mediation to achieve large $A$-terms, and the mechanisms that exist to overcome them. Using these mechanisms, we construct weakly-coupled models of low-scale gauge mediation with extended Higgs-messenger couplings that generate large $A$-terms at the messenger scale and viable $\mu/B_\mu$-terms. Our models are simple, economical, and complete realizations of supersymmetry at the weak scale.}


\begin{document}

\maketitle

\section{Introduction}

The latest results from ATLAS and CMS exclude the Standard Model (SM) Higgs except in the narrow range of $m_h\sim 122-127$ GeV, and show intriguing hints of an excess at $m_h\approx 125$ GeV \cite{ATLAS:2012ae, Chatrchyan:2012tx}. A Standard Model-like Higgs in this range renews the urgency of the hierarchy problem, for which supersymmetry (SUSY) remains the best solution available. Numerous studies have focused on the implications for the MSSM in general (including e.g. \cite{Hall:2011aa, Heinemeyer:2011aa, Arbey:2011ab, Arbey:2011aa, Draper:2011aa, Carena:2011aa, Cao:2012fz,  Christensen:2012ei, Brummer:2012ns}), with the result that  $m_h=125$ GeV in the MSSM translates to a lower bound on a combination of the $A$-terms and the stop mass. For zero stop mixing, the stops must be heavier than $\sim 10$ TeV, and for maximal mixing they must be heavier than $\sim 1$ TeV.
In the former case there is little reason to hope for meaningful signs of supersymmetry at the LHC, and the naturalness problem of the MSSM is greatly exacerbated. We will focus on the latter, more conventional scenario in this paper.

Accurately modeling the Higgs sector is especially challenging in low-scale SUSY-breaking scenarios such as gauge mediation (GMSB; for a review and original references, see e.g.\ \cite{Giudice:1998bp}). There are two reasons for this. The first is omnipresent and pervasive, but is less directly tied to a Higgs at 125 GeV: the well-known $\mu$ - $B_\mu$ problem.  Gauge mediation does not generate the parameters $\mu$ or $B_\mu$ at the messenger scale. Extending gauge mediation to include new interactions in the Higgs sector that generate $\mu$ tends to produce a $B_\mu$-term that is too large for viable electroweak symmetry breaking (EWSB).

The second reason why the Higgs sector is challenging in gauge mediation is a direct consequence of $m_h=125$ GeV. This is the failure of gauge mediation to generate $A$-terms at the messenger scale, in addition to $\mu$ and $B_\mu$. The $A$-terms are instead generated through the renormalization group equations of the MSSM, driven predominantly by the gluino mass. If there is no other source of trilinear soft terms, then in order to generate $A$-terms of sufficient size to explain the Higgs mass, the messenger scale must be extremely high ($M_{\rm mess}\gtrsim 10^{10}$ GeV), and the gluinos must be extremely heavy ($M_{\rm gluino}\gtrsim 3$ TeV) \cite{Draper:2011aa}. Absent additional interactions, this would seem to greatly constrain low-scale supersymmetry breaking.

The purpose of this paper will be to address all of these difficulties in a simple, economical, and calculable setting. To this end, we will construct perturbative spurion-messenger models  that generate the $A$, $\mu$, and $B_\mu$-terms of the right parametric size at the messenger scale. Since vanilla GMSB can generate large $A$-terms through RG evolution from high messenger scales,  in this paper we will focus exclusively on low messenger scales ($M\sim 10^5-10^6$ GeV) where the problem of the Higgs mass is most acute. The models presented here are complete and fully calculable effective theories below the messenger scale; generate all the required couplings of the Higgs sector; and are consistent with collider limits and a Higgs at $m_h=125$ GeV.

The starting point for our model-building is the introduction of marginal superpotential interactions between the Higgses and messengers.\footnote{Alternatives to this would be to consider Higgs-messenger mass mixing \cite{Chacko:2001km, Chacko:2002et,Evans:2011bea, Evans:2012hg,Jelinski:2011xe}; 
interactions between MSSM matter fields and messengers \cite{Dine:1996xk, Giudice:1997ni, Shadmi:2011hs,Jelinski:2011xe}; or perhaps even MSSM matter-messenger mass mixing. Some of these approaches are strongly constrained by precision flavor, for which more intricate model building is required.}
As we will review in the next section, if new Higgs-messenger interactions are introduced, the principal challenge is to generate one-loop $A$-terms at the messenger scale while not generating too large (one-loop) $m_{H}^2$ soft masses. Indeed, just as there is a $\mu$ - $B_\mu$ problem, there is a completely analogous $A$ - $m_{H}^2$ problem. If anything, the  $A$ - $m_{H}^2$ problem is more serious, because $m_{H}^2$ is a singlet under all global symmetries.

Fortunately, the $A$ - $m_H^2$ problem can be solved by adapting a well-known fact: if the sole source of messenger mass is a single SUSY-breaking spurion $X$ as in minimal gauge mediation (MGM)  \cite{Dine:1993yw,Dine:1994vc,Dine:1995ag}, then even in the presence of direct couplings to the messengers, one-loop contributions to scalar mass-squareds vanish to leading order in SUSY-breaking.\footnote{This phenomenon was first noticed in the early literature on gauge mediation   \cite{Dvali:1996cu,Dine:1996xk}. An understanding in terms of the symmetries special to MGM can be found in \cite{Dimopoulos:1997ww,Delgado:2007rz,Giudice:2007ca}.} That is, we may avoid the $A$ - $m_H^2$ problem provided the superpotential takes the form
\beq \label{eq:deltaWintro}
W = X\,\phi_i\cdot\tilde\phi_i + \lambda_{u ij} H_u \cdot\phi_i\cdot\tilde\phi_j
\eeq
with $\langle X\rangle = M+\theta^2 F$ and $i$, $j$ summed over all the messengers (in irreducible representations of $SU(3)\times SU(2)\times U(1)$) of the theory. Here and below, the dots will be used to denote contraction of gauge indices. In order to avoid generating $B_\mu$ at one loop in this model, we must take the analogous coupling for $H_d$ to be zero (or at least extremely small, $\lesssim 10^{-3}$). This can be ensured with an appropriate global symmetry, or by appealing to technical naturalness {\it a la} the SM Yukawa couplings.

As a mechanism for generating large $A$-terms, this was first described in a broader context in \cite{Giudice:1997ni}. Very recently, it was used in \cite{Kang:2012ra} to construct models with an eye specifically towards $m_h = 125$ GeV. Here, we will reanalyze these models with one crucial difference: we will take into account a one-loop, negative, $F/M^2$-suppressed contribution to $m_{H_u}^2$ that was neglected in \cite{Kang:2012ra}. Since $F/M \sim 100$ TeV is fixed by the scale of soft masses, this new contribution is important only when the messenger scale is low -- within a factor of $\sim$ (a few) $\times$ 100 TeV. In this regime, the one-loop contribution opens a qualitatively new region of parameter space for EWSB that is unavailable at higher scales.

Our models for the MSSM (or those of  \cite{Kang:2012ra}) may be viewed as a ``module" for generating large $A$-terms in gauge mediation. One can imagine attaching this module to theories involving a successful solution of the $\mu$ - $B_\mu$ problem without new light degrees of freedom, such as \cite{Giudice:2007ca,Komargodski:2008ax,ourwork}. However, in this paper we explore an alternative and more economical route, one that is all but inexorably suggested by the form of the superpotential (\ref{eq:deltaWintro}). Namely, if we extend the MSSM to the NMSSM, and couple the NMSSM singlet $N$ to the same MGM messengers,  we may simultaneously solve the $\mu$ - $B_\mu$ problem and the $A$ - $m_H^2$ problem! Not only that, but in the NMSSM there is also a need for a negative soft mass and  large trilinears in the singlet potential in order to achieve viable EWSB and avoid ultra-light pseudoscalars \cite{deGouvea:1997cx}.\footnote{For related approaches to this problem, see \cite{Han:1999jc,Delgado:2007rz,Ellwanger:2008py,Morrissey:2008gm}. For a review and references of NMSSM phenomenology, see \cite{Ellwanger:2009dp}.} This is a serious problem in conventional GMSB, since the singlet soft mass-squared $m_N^2$ only arises at three loops and $A$-terms are again small. To a large extent, this has discouraged the pursuit of NMSSM-like models of GMSB, despite the evident suitability of an additional light singlet for addressing the $\mu$ - $B_\mu$ problem. Our extended model with Higgs-messenger and singlet-messenger interactions automatically solves this $A$ - $m_N^2$ problem and reconciles the NMSSM and gauge mediation. Much as before, we find that negative, $F/M^2$-suppressed one-loop contributions to $m_N^2$ open a qualitatively new region of parameter space in the NMSSM models when the messenger scale is low.

So our complete model for $\mu$, $B_\mu$ and large $A_t$ will be:
\beq \label{eq:deltaWintro2}
 W =   X\,(\phi_i\cdot\tilde\phi_i+\varphi_i\cdot\tilde\varphi_i)  + \lambda_{uij} H_u \cdot(\phi_i\cdot\tilde\phi_j +\varphi_i\cdot\tilde\varphi_j) + \lambda_N N ( \phi_i\cdot\tilde\varphi_i) + \lambda N H_u \cdot H_d - {1\over3}\kappa N^3
\eeq
again with $\lambda_d=0$. This superpotential can be made natural under a $U(1)_X\times {\Bbb Z}_3$  symmetry. The doubling of the messenger sector is necessary so that $N$ can couple to a different messenger bilinear than $X$ in order to avoid generating dangerous tadpoles for $N$, which are threatening since it is a gauge singlet \cite{Giudice:1997ni, Delgado:2007rz}.

The addition of singlet-messenger interactions to an NMSSM model in GMSB has been explored previously in \cite{Delgado:2007rz}, for the purpose of generating $\mu$ and $B_\mu$. Here the new ingredient is that we combine it with the $A$-term module in a natural and efficient way in order to generate a suitable mass for the Higgs, together with $\mu$ and $B_\mu$. This combination is far from trivial; as we will see, large $A$-terms place interesting constraints on the NMSSM sector. Indeed, they access a qualitatively different  region of parameter space -- and lower messenger scales --  than were available in \cite{Delgado:2007rz}, and they arguably make it easier to achieve viable EWSB. The interplay of all these issues illustrates the utility in constructing a complete effective theory with the full set of interactions required for viable electroweak symmetry breaking and a sizable Higgs mass.

It bears emphasizing that our philosophy is quite different from the typical approach to the Higgs mass in the NMSSM. Rather than trying to lift the Higgs mass using the NMSSM potential -- an endeavor that is largely incompatible with perturbativity in the Higgs sector up to the GUT scale  -- we instead use the MSSM stop mixing to lift the Higgs mass and only employ the NMSSM to generate $\mu$ and $B_\mu$. Indeed, in this scenario it's easier to generate $m_h \approx 125$ GeV if the NMSSM is in the ``decoupling limit" of $\lambda$, $\kappa\to 0$ with large $\tan\beta$. Otherwise, the NMSSM couplings tend to contribute {\it negatively} to the Higgs mass. In this sense, the Higgs mass and $\mu$, $B_\mu$ have separate origins. Ultimately, however, all the infrared parameters emerge from a common mechanism for generating $A$, $m_H^2$, $\mu$, $B_\mu$ at the messenger scale via interactions with messenger fields.

The low-energy phenomenology of our models is relatively insensitive to the details of the EWSB sector and the choice of messenger representations. To a large extent it resembles that of MGM, due to the key role played by the MGM-like couplings of the messengers to the hidden sector. Concretely, the stops are the lightest colored scalars, and typically the only colored superpartners below 2 TeV. Additional scalars in the EWSB sector are heavy and the Higgs properties are SM-like. Because we are forced to consider larger effective messenger numbers to improve the $A_t/m_{\tilde t}$ ratio, the NLSP is typically the lightest stau. Finally, our exclusive focus on low messenger scales in this paper means that the stau NLSP always decays promptly in the detector.  The most fruitful channels for discovery are likely to be those with leptons and missing energy, and these spectra readily satisfy current LHC limits \cite{Chatrchyan:2012ye}.

It is interesting that we are essentially led to minimality in the messenger sector because of the need to solve the $A$ - $m_H^2$ problem. Since colored superpartners are relatively heavy in MGM, perhaps this explains why we have yet to observe evidence for supersymmetry at the LHC!  Of course, while minimality is appealing from an aesthetic point of view, it is not strictly necessary. The mechanisms we discuss here for generating large $A$-terms without over-large $m_H^2$, and $\mu$ without over-large $B_\mu$, can in principle be added to any general model of gauge mediation \cite{Meade:2008wd}, e.g.\ the model of \cite{Buican:2008ws} which covers the GGM parameter space. This greatly expands the possible phenomenology.

 The outline of our paper is as follows: In section \ref{sec:genprobsol} we present the general problems of gauge mediation in light of a Higgs at 125 GeV, focusing on the challenges of generating a $\mu$-term without an over-large $B_\mu$-term, and likewise large $A$-terms without over-large $m_{H}^2$ soft masses. As we discuss in section \ref{sec:genprobsol}, these problems share a common solution, through the use of minimal gauge mediation and (in case of $\mu$ - $B_\mu$) the NMSSM. We present specific models in section \ref{sec:models}. These include a module for generating large $A$-terms in the MSSM, and a complete theory incorporating $\mu$ and $B_\mu$ in the NMSSM. Various constraints on the models stemming from EWSB and avoidance of tachyons are discussed in detail in section \ref{sec:challenges}, and the spectrum and phenomenology of the models are analyzed in section \ref{sec:results}. We conclude in section \ref{sec:conclusions} with a summary and discussion of future directions. Finally, general formulas for soft masses and a discussion of physics above the messenger scale (i.e.\ Landau poles) are reserved for appendices \ref{app:genmass} and \ref{app:poles}, respectively.

\section{Generalities}
\label{sec:genprobsol}

\subsection{The $\mu$-$B_\mu$ and $A$-$m_H^2$ problems}

A successful theory of supersymmetry breaking should give rise to gaugino masses and scalar soft masses of the same order, as well as $A$-terms and a $B_\mu$-term that are of the same order or smaller.  In addition, supersymmetry breaking should ideally provide a natural origin for the $\mu$-term,
\beq
{\mathcal L} \supset \int  d^2 \theta \, \mu \, H_u H_d ~.
\eeq
Although the $\mu$-term is ostensibly supersymmetric and need not originate from supersymmetry breaking, successful electroweak symmetry breaking requires that the scale of $\mu$ coincide with that of the other soft masses in the Higgs sector, i.e., $\mu^2 \sim m_{\rm soft}^2$. This is the origin of the so-called ``$\mu$ problem''.  A glaring coincidence problem may be avoided only if the $\mu$-term is generated by the same dynamics that breaks supersymmetry.

Gauge mediation gives rise to gaugino masses at one loop and sfermion soft masses-squared at two loops, such that $m_{\tilde g} \sim m_{\tilde f}\sim m_{\rm soft}$ as desired. However, gauge interactions alone do not generate all possible soft terms at similar orders. In particular, in the most general gauge mediation model, $\mu$ and $B_\mu$ are not generated to any order in the gauge couplings, as they are protected by $U(1)_{PQ}$ symmetries which rotate $H_u$ and $H_d$. Meanwhile the $A$-terms are not generated to leading order in the gauge couplings \cite{Meade:2008wd}. They can be generated at higher orders, through the usual MSSM RGEs, but this means that it is quite challenging to make them large enough, especially for $m_h=125$ GeV \cite{Draper:2011aa}. The failure of gauge interactions to generate appropriate contributions to Higgs sector soft parameters suggests that  a viable and complete theory of supersymmetry at the weak scale should include both gauge mediation and additional couplings to the Higgs sector.

This may be arranged in gauge mediation by introducing couplings between the Higgs multiplets and messengers such that, below the scale $M$ of the messengers, the theory contains an effective operator of the form
\beq \label{eq:muterm}
 {\mathcal L} \supset \int d^4 \theta \, \frac{c_\mu}{M} X^\dag H_u H_d + {\rm h.c.}
 \eeq
Here we are working in the spurion limit, where the effects of supersymmetry breaking are encoded by the expectation values $\langle X\rangle = M+\theta^2 F$ and the dynamics of $X$ may be neglected; we also assume the messenger sector is weakly coupled.
The effective operator (\ref{eq:muterm}) leads to a $\mu$-term of the right size in weakly-coupled models provided the coefficient $c_\mu$ arises at one loop.

However, in most models of gauge mediation, whatever Higgs-messenger interactions give rise to (\ref{eq:muterm}) likewise generate an effective operator contributing to the $B_\mu$-term of the form
\beq
{\mathcal L}\supset \int d^4 \theta \, \frac{c_{B_\mu}}{M^2} X^\dag X H_u H_d + {\rm h.c.}
\eeq
at the same loop order. Consequently, one finds $B_\mu / \mu^2 \propto c_{B_\mu} / c_\mu^2 \sim 16 \pi^2 / \lambda_\mu^2 \gg 1$, where $\lambda_\mu$ represents some set of perturbative Higgs-messenger couplings that collectively break the PQ symmetry. This  ``$\mu$ - $B_\mu$ problem'' is a disaster for stable electroweak symmetry breaking, which generally requires $\mu^2 \sim B_\mu$.\footnote{An exception is if $m_{H_d}^2$ is large and positive, in which case the standard EWSB relations in the MSSM allow for $B_\mu \gg \mu^2$. The idea was first proposed in \cite{Dine:1997qj}, together with a viable messenger model involving multiple spurions. More details were later worked out in \cite{Csaki:2008sr,DeSimone:2011va}.} Considerable attention has been devoted to possible solutions to this problem.

Interestingly, an analogous problem -- which has thus far received
much less attention -- arises between $A$-terms and Higgs soft
masses $m_{H_{u,d}}^2$.  An attractive way to generate sizable
$A$-terms aligned with Standard Model Yukawa couplings is to
introduce Higgs-messenger couplings that lead to K\"{a}hler terms of
the form \beq \label{eq:aterm} {\mathcal L}\supset \int d^4 \theta
\, \frac{c_{A_{u,d}}}{M} X H_{u,d}^\dag H_{u,d} + \rm{h.c.} \eeq
Such terms give rise to $A$-terms after substituting
$F_{H_{u,d}}^\dag \to -y_u Q \bar u$ via the MSSM superpotential,
and the $F$-component vev of $X$.\footnote{As we will
discuss in more detail later, these terms also contribute to
$m_{H_{u,d}}^2$, as well as to $B_\mu$ if $\mu$ is present in the
superpotential. 
} The resulting $A$-terms are attractive from the
perspective of flavor physics, since they are naturally aligned with
Standard Model Yukawa couplings. In order for these $A$-terms to
have a meaningful impact on the gauge-mediated soft spectrum,
$c_{A_{u,d}}$ should arise at one loop so that $A \sim m_{\rm
soft}$.

However, in complete analogy with $\mu$ - $B_\mu$, whatever interactions generate (\ref{eq:aterm}) also typically give rise to contributions to Higgs soft masses of the form
\beq
{\mathcal L}\supset \int d^4 \theta \, \frac{c_{m_{u,d}}}{M^2} X^\dag X H_{u,d}^\dag H_{u,d}
\eeq
at the same loop order. If $A$-terms are generated at one loop, as is necessary for them to have any impact on the mass of the Higgs, this implies a one-loop contribution to Higgs soft masses that seriously imperils electroweak symmetry breaking. This ``$A$ - $m_H^2$ problem'' is especially troublesome because the couplings $c_{m_{u,d}}$ are singlets under all possible global symmetries acting on $X$ and $H_{u,d}$, making it difficult to generalize many conventional approaches to the $\mu$ - $B_\mu$ problem.

In order to understand the solution to the $A$ - $m_H^2$ problem, it is useful to reexamine the problem more concretely using the most general perturbative model of messengers coupled to a supersymmetry-breaking spurion \cite{Cheung:2007es}:
\beq \label{eq:Weogm}
W = \lambda_{ij} X \phi_i\cdot\tilde\phi_j + m_{ij}\phi_i\cdot\tilde\phi_j ~.
\eeq
Here as above, $\langle X\rangle = M+\theta^2 F$. If we want one-loop $A$-terms, we should couple the messengers directly to the Higgs fields:\footnote{Note that we could also write down interactions involving singlet messengers $S$ of the form $W \supset S H_u H_d$; we will not consider this option in detail here.}
\beq \label{eq:WAterms}
\delta W = \lambda_{uij} H_u \cdot \phi_i \cdot\tilde\phi_j + \lambda_{dij}H_d\cdot \phi_i\cdot\tilde\phi_j ~.
\eeq
Integrating out the messengers leaves a supersymmetric effective theory for $X$, $H_u$ and $H_d$ described by the effective Kahler potential 
\beq \label{eq:Keff}
K_{eff} =   Z_u(X,X^\dagger)H_u^\dagger H_u+Z_d(X,X^\dagger)H_d^\dagger H_d + ( Z_\mu(X,X^\dagger) H_u H_d + h.c.)  + \dots 
\eeq
where the ellipses denote terms that are higher order in $H_u$ and $H_d$, and $\epsilon$ is a loop-counting parameter. In addition to being functions of $X$ and $X^\dagger$, the wavefunction factors will also depend on other dimensionful parameters. This includes not only $m_{ij}$ from the superpotential, but also a UV cutoff $\Lambda_0$ -- in general the wavefunctions are UV divergent quantities.

The terms in (\ref{eq:Keff}) are responsible for generating $\mu$, $B_\mu$, $m_{H_u}^2$, $m_{H_d}^2$, and $A_u$, $A_d$. Specifically, we have, to leading order in $F/M^2$ and to one-loop order:
\begin{eqnarray} \label{eq:Higgspars} \nonumber
 \mu = F \, \partial_X Z_\mu^{(1)} &&,\qquad B_\mu = |F|^2 \, \partial_X\partial_{X^\dagger} Z_\mu^{(1)} \\
 A_{u,d} = F \, \partial_X  Z_{u,d}^{(1)} &&,\qquad m_{H_{u,d}}^2 = |F|^2 \, \partial_X\partial_{X^\dagger}  Z_{u,d}^{(1)}
\end{eqnarray} 
where all the $X$ derivatives are evaluated at $X=M$. If the wavefunctions are completely general functions of $X$, $X^\dagger$, $m_{ij}$ and $\Lambda_0$, typically nothing will cause their mixed second derivatives to disappear, and so nothing prevents $B_\mu$, $m_{H_u}^2$, and $m_{H_d}^2$ from appearing at one loop at leading order in $F/M^2$. Although one might hope to forbid certain terms from appearing in the wavefunction factors using appropriate global symmetries, the soft terms $m_{H_{u,d}}^2$ are especially dangerous, as they are neutral under all global symmetries.

In summary, we see that if $\mu$ is generated at one loop, then $B_\mu$ also tends to be generated at one loop; there is nothing in the form of (\ref{eq:Higgspars}) that distinguishes the loop counting of the two parameters. Similarly, if $A_{u,d}$ is generated at one loop, then $m_{H_{u,d}}^2$ tends to be generated at one loop. So just as there is a $\mu$ - $B_\mu$ problem, there is an $A$ - $m_{H}^2$ problem.

\subsection{A general mechanism for a solution}
\label{seq:generalmech}

As is well known (see especially the discussion in \cite{Giudice:2007ca}), there is one special case for which the mixed derivatives of the wavefunction factors will vanish: minimal gauge mediation \cite{Dine:1993yw,Dine:1994vc,Dine:1995ag}, for which the only source of mass in the messenger sector is the vev of $X$.  In that case, the superpotential is constrained to take the form $W = \lambda_i X \phi_i\tilde\phi_i$. In fact, this model is further special: it is endowed with an $R$-symmetry under which $R(\phi_i)=R(\tilde\phi_i)=0$ and $R(H_u)=R(H_d)=R(X)=2$. This $R$-symmetry is broken only by the lowest-component-vev of $X$. Then by a combination of dimensional analysis and the  $R$-symmetry, the wavefunction renormalization factors are constrained to take the form
\beq \label{eq:Zmgm}
Z_{u,d} = f \left({X^\dagger X\over \Lambda_0^2} \right),\qquad Z_\mu = \left({X^\dagger \over X}\right) g \left({X^\dagger X\over\Lambda_0^2} \right) ~.
\eeq
At one loop, we can have at most a logarithmic divergence by power counting. So symmetries and dimensional analysis imply $Z_{u,d}^{(1)} = c_{u,d} \lambda_{u,d}^2 \log X^\dagger X /\Lambda_0^2$, in which case the one-loop contributions to $m_{H_{u,d}}^2$ vanish! Of course, we emphasize that this approach only captures the leading-order effects in $F/M^2$, so that there may be nonzero one-loop soft masses suppressed by powers of $F/M^2$ that are not problematically large.

Meanwhile, we see from (\ref{eq:Zmgm}) that $B_\mu$ does not vanish in general at one loop, and typically must be forbidden by imposing additional symmetries. In fact, if $g$ is a nontrivial function, $\mu$ and $B_\mu$ can in general be UV sensitive. These problems may be avoided if $\lambda_d = 0$ in (\ref{eq:WAterms}), in which case there is an additional PQ symmetry; the one-loop contribution  $Z^{(1)}_\mu  \propto \lambda_u \lambda_d$ vanishes and neither $\mu$ nor $B_\mu$ arise at this order. Since we wish to generate $A$-terms without exacerbating the $\mu$ - $B_\mu$ problem, in what follows we will exploit this case and take $\lambda_d = 0$. This choice is technically natural, and may be enforced by a global symmetry distinguishing $H_u$ and $H_d$.

Although this approach leads to sizable $A$-terms and solves the $A$ - $m_{H}^2$ problem, it does not explain the origin of $\mu$ and $B_\mu$. While it is possible to address the problem by supplementing the messenger sector with additional interactions and symmetries, there exists a far more economical route. Namely, if we extend the MSSM by a single light singlet field $N$, and couple $N$ to the same MGM messengers that $H_u$ couples to, then we can simultaneously generate $\mu$, $B_\mu$ and $A_t$! In this paper, we will focus on the simplest scenario, namely the ${\Bbb Z}_3$ symmetric NMSSM:
\beq
W \supset \lambda N H_u \cdot H_d - {1\over3}\kappa N^3  ~.
\eeq
As is well known, the main obstacle to marrying the NMSSM and gauge mediation is that a viable vacuum requires a sufficiently large, negative soft mass $m_N^2$ at the weak scale, as well as sizable trilinear couplings $A_\lambda$, $A_\kappa$ -- but pure gauge mediation does not generate any of these quantities at the messenger scale \cite{deGouvea:1997cx}. So by the same logic as before, one is confronted with an $A$ - $m_N^2$ problem in the NMSSM. But again, the same logic tells us that there is a uniform solution of all of these problems -- $A$ - $m_N^2$, $A$ - $m_H^2$, and $\mu$ - $B_\mu$ -- via the MGM-messenger mechanism described above!

\subsection{The little $A$-$m_H^2$ problem}
\label{subsec:littleamh}

Thus far our discussion of viable spectra has focused on the loop order at which various soft parameters arise. While a necessary constraint, it is not sufficient on its own to guarantee successful electroweak symmetry breaking; even if they are all the same size, the Higgs sector soft parameters must satisfy various inequalities in order to ensure a nontrivial vacuum. In particular, the soft masses $m^2_{H_u}$ and $m_{\tilde t}^2$ receive large corrections at the messenger scale from the Higgs-messenger coupling $\lambda_u$, such that radiative electroweak symmetry breaking may no longer be taken for granted. The contributions to $m_{H_u}^2$ are highly generic and particularly troublesome. As mentioned above, whenever $A$-terms arise via K\"{a}hler operators of the form (\ref{eq:aterm}), there is an irreducible contribution  to $m_{H_{u,d}}^2$  given by $A_{u,d}^2$. These arise from putting the auxiliary fields to their equations of motion, e.g.: 
\begin{equation}
\label{eq:littleAmH}
-V \supset F_{H_u}^\dagger F_{H_u} +\left( A_u H_u F_{H_u}^\dagger + c.c.\right) \to -A_u^2 H_u^\dagger H_u
\end{equation}
Although this increase in $m_{H_u}^2$ does not necessarily spoil electroweak symmetry breaking, it greatly enhances the degree to which the model is tuned. Thus even when the loop-level $A$ - $m_H^2$ problem is solved, there is a remnant ``little $A$ - $m_H^2$'' problem that is universal in models where the $A$-terms originate from K\"{a}hler operators such as (\ref{eq:aterm}).  


The consequences for EWSB depend on the specific choice of messenger representations and couplings. We will first present  general models for the MSSM and the NMSSM, and reserve a detailed discussion of electroweak symmetry breaking for section \ref{sec:challenges}.

\section{Models}
\label{sec:models}

\subsection{Warmup: an MSSM module for large $A$-terms}
\label{subsec:mssmmod}

Let us now analyze an explicit model with the features discussed above. This model was constructed recently in \cite{Kang:2012ra}, motivated by $m_h=125$ GeV. Our analysis in this paper will differ crucially in the treatment of one-loop soft masses. As we will show, these can have profound effects on the model at low messenger scales.

Consider a theory with messengers $\phi_i$, $\tilde \phi_i$ in vector-like irreps of  $SU(3)\times SU(2)\times U(1)$; a SUSY-breaking spurion $X$ with $\langle X \rangle = M + F \theta^2$; and superpotential interactions
\beq \label{eq:gensuper}
W = X \, \phi_i\cdot\tilde\phi_i+\lambda_u \,H_u\cdot \phi_1\cdot\tilde\phi_2  + y_t \,H_u \cdot Q \cdot U  + \mu \, H_u \cdot H_d + \dots
\eeq
where the ellipses denote other MSSM interactions that are irrelevant for our purposes. Here we are making a number of simplifying assumptions: first, we are assuming that there is only one combination of the messengers, $\phi_1\cdot\tilde\phi_2$, that can be combined with $H_u$ to make a gauge singlet. The generalization to multiple such couplings is straightforward. Second, at this stage we are interested in generating $A$-terms at one loop, rather than explaining the origin of $\mu$ and $B_\mu$, and so we will allow for arbitrary $\mu$ and $B_\mu$.\footnote{As noted earlier, the Higgs-messenger interactions contribute to $B_\mu$ given a supersymmetric $\mu$-term, but we allow arbitrary additional contributions to satisfy EWSB.} In the next subsection, we will extend the model to also generate these parameters. Finally, we are only including the top Yukawa explicitly in (\ref{eq:gensuper}), because its large size means that it will play a role in the later analysis.\footnote{The bottom and tau yukawas are unimportant even at large $\tan \beta$ because our Higgs-messenger couplings only involve $H_u$.}

The interactions (\ref{eq:gensuper}) comprise the most general renormalizable superpotential consistent with the SM gauge symmetry, together with messenger number (to forbid messenger-matter mixing), and a global $U(1)_X$ symmetry under which the fields carry the following charges:
\beq\label{eq:mssmcharges}
q_X(X,\phi_i,\tilde\phi_i,H_u,H_d) = (1,-1/2,-1/2,1,-1) ~.
\eeq
 (Charges of MSSM matter fields can always be chosen such that the usual Yukawa terms are allowed.) Messenger number forbids mixing with matter multiplets and renders the lightest messenger stable, though this may be readily broken by higher-dimensional operators \cite{Dimopoulos:1996gy}.

We may readily extend this model to include $N_{mess}$ flavors of messengers, so in general we consider messengers $\phi_{if}$, $\tilde\phi_{if}$ with $f=1,\dots N_{mess}$. To avoid a proliferation of couplings, we will impose a $U(N_{mess})$ flavor symmetry.

The superpotential interactions (\ref{eq:gensuper}) give rise to both conventional gauge-mediated soft masses and new contributions to $ m_{H_u}^2$, $ m_Q^2$, $ m_U^2$, and $A_{t}$ due to the direct Higgs-messenger interaction. The latter are given by (see appendix \ref{app:genmass} for the derivation and references):
\begin{eqnarray}\label{eq:mhusqfinal}
 && \delta m_{H_u}^2 =- d_H \frac{\alpha_{\lambda_u}}{12 \pi} h(\Lambda/M) \left({\Lambda\over M}\right)^2\Lambda^2+ \left( d_H(d_H+3) {\alpha_{\lambda_u}^2\over 16\pi^2}  - d_H C_r{\alpha_r \alpha_{\lambda_u}\over 8\pi^2} \right)\Lambda^2 \\
\label{eq:mqsq}  & &\delta m_{Q}^2  = -d_H{ \alpha_t\alpha_{\lambda_u}\over 16\pi^2} \Lambda^2 \\
\label{eq:musq}  && \delta m_{U}^2  = -d_H {\alpha_t\alpha_{\lambda_u}  \over 8\pi^2} \Lambda^2 \\
 \label{eq:Atfinal} & &A_{t} = -d_H {\alpha_{\lambda_u}\over 4\pi} \Lambda
\end{eqnarray}
Here we have introduced
\begin{equation}
\label{eq:Lambdadef}
\Lambda\equiv F/M
\end{equation}
Also, $d_H$ counts the total number of fields coupled to $H$ through $\lambda_u$;  and $C_r=c_r^{H_u}+c_r^{\phi_1}+c_r^{\tilde\phi_2}$ is the sum of quadratic Casimirs of the fields which participate in the Higgs-messenger-messenger Yukawa coupling.
(Concrete examples of $d_H$ and $C_r$ will be given in section \ref{sec:explicitmodel}.) The little $A$ - $m_H^2$ problem is manifest in the second term of (\ref{eq:mhusqfinal}), specifically in the contribution proportional to $d_H^2$. 

The first term in $\delta m_{H_u}^2$ is the $\Lambda/M$-suppressed one-loop contribution to $m_{H_u}^2$ which cannot be eliminated by the MGM mechanism described in the previous section. The function $h(x)$ is given by:
\beq \label{eq:h}
h(x) = \frac{3\Big( (x-2) \log(1-x) - (x+2) \log(1+x) \Big)}{x^4} = 1 + \frac{4x^2}{5} + \dots
\eeq
and is such that the one-loop contribution to $m_{H_u}^2$ is always strictly negative. This effect was neglected in \cite{Kang:2012ra}, and it will be crucial for the discussion in section \ref{secEWSB}, when we analyze the viability of these models from the perspective of EWSB. There are, of course, additional $\Lambda/M$-suppressed contributions to all the other soft masses \cite{Dimopoulos:1996gy, Martin:1996zb}, but these are always subdominant. The $\Lambda/M$-suppressed contribution to $m_{H_u}^2$ (and only $m_{H_u}^2$) is parametrically important because it first arises at one loop.

Once the  number and type of messenger representations are specified, the dimensionless parameter space of the MSSM module for large $A$-terms consists solely of $(\Lambda/M, \lambda_u)$. In addition, there is one dimensionful parameter $\Lambda$ that sets the overall scale of the soft masses. Since we are interested in a particular Higgs mass,  $m_h = 125$ GeV, this completely fixes $\Lambda$, given a choice of $(\Lambda/M,\lambda_u)$. We will explore this parameter space in detail in sections \ref{sec:challenges} and \ref{sec:results}.

We emphasize that the addition of these Higgs-messenger interactions to the MSSM is essentially modular.  It leaves unaltered (and unaddressed) whatever physics generates $\mu$ and $B_\mu$, and may be incorporated into a variety of solutions to the  $\mu$ - $B_\mu$ problem. In general, new interactions that generate $\mu$ and $B_\mu$ also contribute to $m_{H_u}^2$, often at two loops. The sign of these contributions to $m_{H_u}^2$ depend on the details of the model, and may either increase or decrease the value of $m_{H_u}^2$ at the messenger scale. Scenarios in which the new contributions are negative  \cite{Komargodski:2008ax, ourwork} will improve the prospects for (radiative)  electroweak symmetry breaking.

\subsection{The complete NMSSM model for $A$, $\mu$, and $B_\mu$}
\label{sec:nmssm}

While the coupling of messengers to $H_u$ in the MSSM provides an avenue for generating $A$-terms at one loop, it does not explain the origin of $\mu$ and $B_\mu$. Indeed, had we allowed analogous couplings to $H_d$, we would have generated both a $\mu$-term and a $B_\mu$-term at one loop, which would have been disastrous for EWSB. This suggests another source is required for the $\mu$ and $B_\mu$-terms. A natural possibility is the addition of a gauge singlet superfield $N$, which may be coupled to messengers much like $H_u$ \cite{Giudice:1997ni, Delgado:2007rz}.

As discussed in the previous section, the addition of a light gauge singlet superfield raises the usual challenges of generating suitable $A$-terms and $m_{N}^2$ in the singlet sector. This is again solved by the same MGM mechanism.\footnote{The same approach was also used recently in \cite{Donkin:2012yn}. However the EWSB mechanism in this paper is different from ours, as they require the ``lopsided" hierarchy $\mu^2\sim m_{H_u}^2\ll B_\mu\ll m^2_{H_d}$ 
\cite{Dine:1997qj,Csaki:2008sr,DeSimone:2011va}.}
However, the new challenge is that $N$, being a gauge singlet, can potentially mix with $X$, leading to dangerous tadpole terms for $N$ \cite{Giudice:1997ni, Delgado:2007rz}. To forbid these, it suffices to extend the $U(1)_X$ symmetry of (\ref{eq:mssmcharges}) to include $N$ so that\footnote{Note that in \cite{Giudice:1997ni, Delgado:2007rz}, a ${\Bbb Z}_3$ symmetry was invoked for this purpose. However, their ${\Bbb Z}_3$ symmetry is neither sufficient nor necessary for  obtaining a viable model. In particular, it does not forbid direct EOGM \cite{Cheung:2007es} mass terms for the messengers.}
\beq\label{eq:nmssmcharges}
q_X(N) = 0 ~.
\eeq
By itself though, this would forbid any coupling between $N$ and the messengers. So we will follow the approach of \cite{Giudice:1997ni, Delgado:2007rz} and double the messenger sector, $\phi_i\to \phi_i$, $\varphi_i$, using the freedom to assign different $U(1)_X$ charges to $\phi_i$ and $\varphi_i$. For instance, we can take the following charge assignment:
\beq\label{eq:nmssmchargesii}
q_X(X,\phi,\tilde\phi,\varphi,\tilde\varphi,H_u,H_d,N) = (1,0,-1,-1,0,1,-1,0) ~.
\eeq
With this $U(1)_X$ symmetry, we can have a viable model with the superpotential
\beq\label{eq:WNMSSM}
W = X(\phi_i\cdot\tilde\phi_i+\varphi_i\cdot\tilde\varphi_i)  + \lambda_{u}H_u\cdot(\phi_1\cdot\tilde\phi_2+\varphi_1\cdot\tilde\varphi_2) + \lambda_N N \phi_i\cdot\tilde\varphi_i + \lambda N H_u \cdot H_d  + \dots
\eeq
where the ellipses again denote other MSSM interactions that are irrelevant for our purposes.

At this point $N$ is a total singlet and so its interactions can be fully general, for instance those discussed in \cite{Ellwanger:2009dp}. Of course, a total singlet with arbitrary interactions is disastrous for many reasons, including the possible reintroduction of UV divergences; typically some set of symmetries must be imposed to ensure that the theory is well-behaved.  In this paper, we will for simplicity focus on the usual ${\Bbb Z}_3$-symmetric NMSSM. Because of the $N$-messenger-messenger couplings, this ${\Bbb Z}_3$ must be extended to act on the messengers as well. A consistent charge assignment is:
\begin{equation}
\label{nmssmchargesZ3}
{\Bbb Z}_3(X,\phi_i,\tilde\phi_i,\varphi_i,\tilde\varphi_i,H_u,H_d,N) = (0, 1,2,2, 1,   0, 2,   1) ~.
\end{equation}
Then the most general superpotential consistent with our choice of $U(1)_X \times {\Bbb Z}_3$ symmetry is:
\beq\label{eq:WNMSSMZ3}
W = X(\phi_i\cdot\tilde\phi_i+\varphi_i\cdot\tilde\varphi_i)  + \lambda_{u}H_u\cdot(\phi_1\cdot\tilde\phi_2+\varphi_1\cdot\tilde\varphi_2) + \lambda_N N \phi_i\cdot\tilde\varphi_i + \lambda N H_u \cdot H_d  - {1\over 3}\kappa N^3 + y_t H_u \cdot Q  \cdot U + \dots
\eeq
This is our complete model of supersymmetry at the messenger scale, which will give rise to all the superpartner masses, large $A_t$, $\mu$, and $B_\mu$ after the messengers are integrated out. Note that the model without the $H_u$-messenger interaction was studied in \cite{Delgado:2007rz}. We will see that adding this interaction and requiring large $A_t$ for $m_h=125$ GeV qualitatively changes the model and actually improves its viability.

As we will show in appendix \ref{app:genmass}, the contributions to the soft masses from the NMSSM couplings are given by (these should be added to the standard gauge mediation terms and those in (\ref{eq:mhusqfinal})-(\ref{eq:Atfinal})) :
\begin{eqnarray}\label{msqfinalnmssm}
&& \delta m_{H_u}^2 = \Bigg(
 d_H {\alpha_{\lambda_N}\alpha_{\lambda_u}\over 16\pi^2}
-d_N { \alpha_\lambda  \alpha_{\lambda_N} \over16\pi^2}
 \Bigg)\Lambda^2
 \\
 && \delta m_{H_d}^2 = \Bigg(
  -d_H { \alpha_\lambda \alpha_{\lambda_u}\over16\pi^2}
  -d_N{  \alpha_\lambda  \alpha_{\lambda_N} \over16\pi^2} \Bigg)\Lambda^2
  \\
  \label{eq:mNsq}
 && m_N^2 = -d_N \frac{\alpha_{\lambda_N}}{12 \pi} h(\Lambda/M)\left({\Lambda\over M}\right)^2 \Lambda^2+ \Bigg(
d_N (d_N+2) { \alpha_{\lambda_N}^2\over16\pi^2}
  - d_N {\alpha_{\lambda_N}\alpha_\kappa \over 4\pi^2} -d_H { \alpha_\lambda\alpha_{\lambda_u} \over 8\pi^2}\\ \nonumber
  &&\qquad\qquad  -d_{N}^{i i}c^i_r  {\alpha_{\lambda_N}  \alpha_r\over 4\pi^2}  +\left(d_{N}^{11}d_1^{H2}+d_{N}^{22}d_2^{H1} \right) {\alpha_{\lambda_u}  \alpha_{\lambda_N}\over 16\pi^2}
\Bigg) \Lambda^2
 \\
 && \delta m_{Q}^2  = \delta m_U^2 =\delta A_t = 0 \\
 \label{alambda}
 && A_\lambda = - \left( d_{H} {\alpha_{\lambda_{u}} \over 4\pi}+ d_N {\alpha_{\lambda_N}\over 4\pi}\right)\Lambda\\
\label{akappa}
&& A_\kappa  = -3 d_N{ \alpha_{\lambda_N}\over 4\pi} \Lambda
 \end{eqnarray}
where again, $\Lambda\equiv F/M$; $d_H$ is as above;  and $d_N$ similarly counts the total number of fields coupling to $N$ via $\lambda_N$. We also have $d_N^{ii}$ counting the number of fields of type $\phi_i$ (or $\varphi_i$) coupling to $N$ (and so $d_N = \sum_i d_N^{ii}$). The numbers $d_{1}^{H2}$ and $d_2^{H1}$ count the number of fields coupling to $\phi_1$ and $\phi_2$ respectively through the $\lambda_u$ Yukawa coupling. Finally, $c_r^i$ is the quadratic casimir of $\phi_i$ in the $r$th gauge group. Concrete examples of all of these parameters are given in section \ref{sec:explicitmodel}.

The full NMSSM model introduces three new parameters $(\lambda, \kappa, \lambda_N)$ relative to the MSSM module; as we will discuss in detail in section \ref{sec:EWSBNMSSM}, EWSB fixes two of the extra parameters (say $\kappa$ and $\lambda_N$) in terms of the third ($\lambda$) and the other Higgs sector parameters.   So for fixed messenger content, the full parameter space of the theory is $\Lambda$, $M$, and $\lambda_u$ (the MSSM parameters), plus $\lambda$. Restricting our attention to $m_h = 125$ GeV, the full parameter space may be specified by $\Lambda/M, \lambda_u,$ and $\lambda$. As with the MSSM, it is no longer obvious that EWSB is viable due to new contributions to the Higgs soft mass; we reserve a detailed study for section \ref{sec:challenges}.

\subsection{Examples}
\label{sec:explicitmodel}

Here we will illustrate the use of the general formulas above with two specific examples for messenger representations. These examples effectively exhaust the possibilities for low-scale GMSB consistent with genuine perturbative gauge coupling unification. The only additional possibility which we are not considering here is a ${\bf 24}$ of $SU(5)$, as it is incompatible with the messenger doubling needed for the complete NMSSM model.

\begin{itemize}

\item The first example is where the messengers fill out ${\bf 5}\oplus{\bf\overline{5}}$ representations of $SU(5)$ (plus the necessary gauge singlets to form the $H_u$ Yukawa coupling). In more detail, we take the $SU(3)\times SU(2)\times U(1)$ representations of the fields in (\ref{eq:gensuper}) and (\ref{eq:WNMSSMZ3}) to be
\beq\label{eq:fivemodelfields}
(\phi_1,\phi_2,\phi_3)=\left( ({\bf 1},{\bf 1},0),  ({\bf 1},{\bf  2},1/2), ({\bf 3},{\bf 1},-1/3)\right)
\eeq
Note that the first field is a pure gauge singlet, and the third is purely a spectator in the MSSM case from the perspective of generating $A$-terms, serving only to complete the GUT multiplet and communicate SUSY breaking to colored fields via gauge interactions.

In this model, the quantities needed to fully specify the MSSM and NMSSM soft masses are given by
\begin{eqnarray}
\label{eq:dcfive}
 && d_H = N_{mess},  \quad d_1^{H2} =2,\quad d_2^{H1} =1\nonumber  \\
 && d_N=3N_{mess},\quad d_N^{11}={1\over2}N_{mess},\quad d_N^{22}=N_{mess},\quad d_N^{33}={3\over2}N_{mess}  \\
 && c_r^1 = (0,0,0), \quad c_r^2 = (0,3/4,3/20),\quad c_r^3=(4/3,0,1/15)\nonumber
\end{eqnarray}
Note that for NMSSM models, $N_{mess}$ must be even due to the doubling of the messenger sector.

\item Our second example is where the messengers fill out a ${\bf 10}\oplus{\bf\overline{10}}$ of $SU(5)$. So here we also have three messengers (up to an overall multiplicity $N_{mess}$)
\beq\label{eq:tenmodelfields}
(\phi_1,\phi_2,\phi_3)=\left( ({\bf 3},{\bf  1},2/3),\, ({\bf 3},{\bf 2},1/6), \, ({\bf 1},{\bf 1},1)\right)
\eeq
Note that $\phi$ is actually {\it not} a ${\bf 10}$ of $SU(5)$. Rather, $\phi$ plus the conjugate $\tilde\phi$ fill out a ${\bf 10}\oplus{\bf\overline{10}}$ of $SU(5)$; this choice is merely a notational convenience that makes manifest the global symmetries of the theory. The key difference relative to  the ${\bf 5 + \overline{5}}$ case is that the messengers coupling to the Higgs are now charged under $SU(3)$. As was shown in \cite{Kang:2012ra}, this has interesting consequences for the viable parameter space of the theory due to negative contributions to $m^2_{H_u}$ proportional to $\sim\alpha_3\alpha_{\lambda_u}$ in (\ref{eq:mhusqfinal}).

In this model, for $N_{mess}$ pairs of ${\bf 10}\oplus {\bf\overline{ 10}}$ messengers, the necessary coefficients for soft parameters are given by
\begin{eqnarray}
\label{dcten}
 && d_H = 3N_{mess}, \quad d_1^{H2} =2,\quad d_2^{H1} =1\nonumber\\
 && d_N=5N_{mess},\quad d_N^{11}={3\over2}N_{mess},\quad d_N^{22}=3N_{mess},\quad d_N^{33}={1\over2}N_{mess}  \\
 && c_r^1 = (4/3,0,4/15), \quad c_r^2 = (4/3,3/4,1/60),\quad c_r^3=(0,0,3/5)\nonumber
\end{eqnarray}
Again, for NMSSM models, $N_{mess}$ must be even.

\end{itemize}

\section{EWSB and Other Constraints}
\label{sec:challenges}

It is clear thus far that introducing additional interactions to generate sizable $A$-terms has repercussions for the rest of the theory via new contributions to various soft masses. These new contributions are typically quite large (since $A$-terms are large) and may significantly alter the vacuum structure of the theory relative to MGM. Requiring a viable vacuum in which electroweak symmetry is broken but various other Standard Model symmetries are preserved leads to nontrivial constraints on the space of UV parameters. In particular, the challenges of guaranteeing EWSB while avoiding charge- and color-breaking minima are more acute than in MGM and favor particular values of scales and couplings in the effective theory.  In this section we will discuss the qualitative challenges imposed by a viable vacuum and study several benchmark points that exemplify the effects on parameter space.  These effects will be further manifest in section \ref{sec:results}, where we perform a comprehensive numerical study on the viable parameter space for explicit models.

\subsection{EWSB in the MSSM with large $A$-terms}
\label{secEWSB}

In the MSSM, minimizing the Higgs potential leads to the relations
\begin{eqnarray}
\label{eq:vacuummin1}
\mu^2 &=& \frac{m_{H_d}^2 - m_{H_u}^2 t_\beta^2 }{t_\beta^2 - 1} - \frac{1}{2} m_Z^2 \\
\label{eq:vacuummin2}
s_{2 \beta} &=& \frac{2 B_\mu}{m_{H_d}^2 + m_{H_u}^2 + 2 \mu^2}
\end{eqnarray}
between soft parameters and the observables $m_Z$ and $\tan \beta$ at the scale of EWSB. (In all the numerical calculations that follow, we will use $\tan\beta=10$.)
These conditions are satisfied at the minimum of the potential, but alone are not sufficient to guarantee a viable vacuum. Rather, the soft parameters must also satisfy the inequalities
\begin{eqnarray}
\label{eq:ineq1}
 & &  B_\mu < |\mu|^2 + {1\over2}(m_{H_u}^2 + m_{H_d}^2)\\
  \label{eq:ineq2}
 & &  B_\mu^2 > (|\mu|^2 + m_{H_u}^2)(|\mu|^2 + m_{H_d}^2)
  \end{eqnarray}
which correspond to the requirements that the quadratic part of the scalar potential is positive along $D$-flat directions and that the origin is not a stable minimum (in which case the vacuum could preserve electroweak symmetry), respectively.

These conditions are most readily satisfied if  $m_{H_u}^2$ is negative at the weak scale. Indeed, renormalization group evolution of soft parameters down to the weak scale typically ensures that this is the case. The most salient feature of the RG evolution of $m_{H_u}^2$ is the negative contribution coming from the large top quark Yukawa, proportional to the third generation soft masses $m_{Q_3}^2, m_{u_3}^2$ and the top $A$-term.  With high messenger scales, the large logarithm is usually sufficient to guarantee that these contributions drive $m_{H_u}^2$ negative. With low messenger scales, as we are focusing on in this paper, the logarithm is not large, so $m_{H_u}^2$ must be negative for other reasons. In GMSB, the soft masses of colored scalars are substantially larger than the messenger-scale soft mass for $H_u$ due to the size of the QCD gauge coupling and the $SU(3)$ Casimir; this suffices to drive $m_{H_u}^2$ negative in only a few decades of running. So even though $m_{H_u}^2$ is positive at the messenger scale, radiative effects drive it negative before the weak scale. Thus radiative electroweak symmetry breaking is a robust prediction of minimal GMSB.

In {\it any} model (not just GMSB) with sizable contributions to Higgs $A$-terms at the messenger scale,  the success of radiative EWSB is greatly endangered. Even if the $A$ - $m_H^2$ problem is solved at one loop, it is in general impossible to suppress the two-loop contributions to $m_{H_u}^2$.  Since generating $m_h\sim 125$ GeV requires $A$-terms at least as large as the stop masses, and since the $A$-terms and $m_{H_u}^2$ have a common origin, it is generally the case that $m_{H_u}^2\sim m_{stop}^2$. But then RG contributions from third generation soft masses are no longer sufficient to drive $m_{H_u}^2$ negative when running from a low scale. While $A$-terms also act to drive $m_{H_u}^2$ negative, they are not parametrically larger than $m_{H_u}^2$ itself. So the success of radiative EWSB now depends sensitively on the messenger scale.  All of these features are illustrated concretely in (\ref{eq:mhusqfinal})-(\ref{eq:Atfinal}), but the problem of EWSB is highly generic.

\begin{figure}[t]
\includegraphics[width=6in]{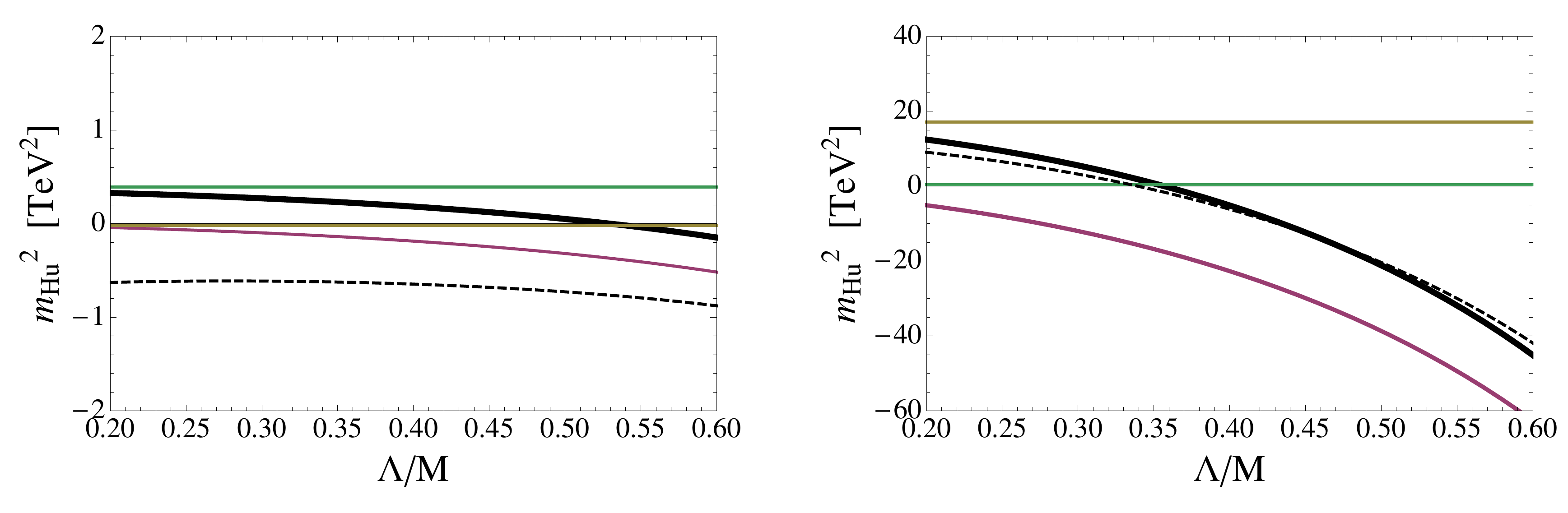}
\caption{Plot of $m_{H_u}^2$ vs $\Lambda/M$, for $\Lambda=110$ TeV,
$N_{mess}=4$, and $\lambda_u=0.1$ (left panel) and  $\lambda_u=1.1$
(right panel).  The total value of $m^2_{H_u}$ in the MSSM  at the
messenger scale (see (\ref{eq:gensuper})) is shown in black. The
green, red and yellow lines indicate the usual GMSB contribution,
the one-loop contribution from $\lambda_u$, and the two- loop
contribution from $\lambda_u$, respectively. Finally, the dashed
black line represents the value of $m^2_{H_u}$ RG evolved down to
$M_{SUSY}$. EWSB at large $\tan\beta$ requires $m_{H_u}^2<0$ at the
weak scale. We see that for large $\lambda_u$, EWSB is achieved for
sufficiently large $\Lambda/M$ due to the negative 1-loop
contribution. For small values of $\lambda_u$ EWSB is achieved
radiatively, as in MGM.\label{fig:mhuVsLambdaOverM}}
\end{figure}

\begin{figure}[t]
 \begin{center}
\includegraphics[width=4.5in]{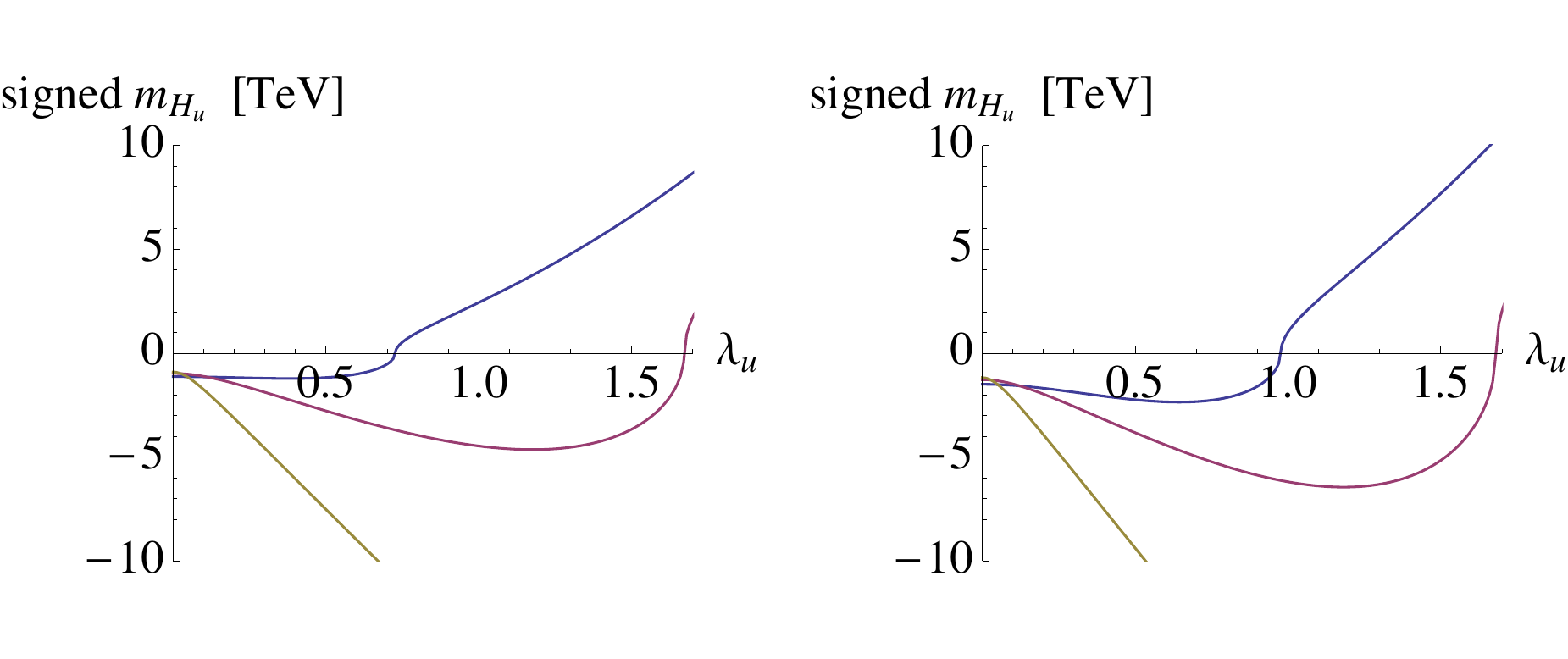}
\end{center}
\caption{Plot of the weak-scale signed mass $m_{H_u}^2/\sqrt{|m_{H_u}^2|}$ vs $\lambda_u$, for the ${\bf 5}\oplus {\bf\overline{5}}$ model  with  $N_{mess}=4$ (left) and the ${\bf 10}\oplus{\bf\overline{10}}$ model with  $N_{mess}=2$ (right). In both plots, $\Lambda=110$ TeV. The blue, red and yellow curves correspond to $\Lambda/M=0.1$, 0.5 and 0.9 respectively. As we will see, $m_h\sim 125$ GeV requires $\lambda_u\sim 1$ for reasonable stop masses. We see that in the ${\bf 5}\oplus{\bf\overline{5}}$ model, $m_{H_u}^2$ becomes positive well below $\lambda_u \sim 1$ for $\Lambda/M=0.1$, but not for $\Lambda/M=0.5$ or 0.9. But in the ${\bf 10}\oplus{\bf\overline{10}}$ model, even $\Lambda/M=0.1$ is possible, because $m_{H_u}^2$ is receiving an additional negative contribution from the colored messengers. } \label{fig:mhuvslambdau}
\end{figure}

However, all is not lost. In the context of our models, we identify two possibilities for rescuing EWSB:

\begin{itemize}

\item If the messenger scale is low ($M \lesssim 10^6$ GeV), then the negative, $\Lambda/M$-suppressed one-loop contribution to $m_{H_u}^2$ in (\ref{eq:mhusqfinal}) may be competitive with the unsuppressed two-loop contribution. Partial or complete cancellation between the two terms of different loop order may diminish the value of $m_{H_u}^2$ at the messenger scale or render it negative. The effect is illustrated in figure \ref{fig:mhuVsLambdaOverM}. The 1-loop contribution was neglected in \cite{Kang:2012ra}, but we will see that it significantly influences theories with low messenger scales; it will also play an additional key role when we turn to the NMSSM.

 \item Alternatively, there can be a significant reduction of the two-loop contribution itself,  if the gauge contribution in (\ref{eq:mhusqfinal}) is large enough to partially or wholly cancel the Yukawa contribution \cite{Kang:2012ra}. Since obtaining the physical Higgs mass through stop mixing requires $\lambda_u \gtrsim g_3$, among the Standard Model gauge couplings only $g_3$ is large enough to result in meaningful cancellation.\footnote{Also considered in \cite{Kang:2012ra} is the possibility that the messengers are charged under a strong hidden sector gauge group.} Thus if any of the messengers $\phi_1$, $\tilde \phi_1$, $\phi_2$, $\tilde\phi_2$ are charged under $SU(3)_C$, the two-loop contributions to $m_{H_u}^2$ may largely cancel among themselves for arbitrary messenger scale. Note that this is impossible to arrange when the messengers transform as complete ${\bf 5 + \bar{5}}$ multiplets, but may occur if they transform as higher-rank representations such as ${\bf 10 + \overline{10}}$. In this case, $m_{H_u}^2$ is still typically positive at the messenger scale, but is small enough to be driven negative by radiative effects before the weak scale. This effect is illustrated in fig. \ref{fig:mhuvslambdau}.

\end{itemize}

\subsection{EWSB in the NMSSM}
\label{sec:EWSBNMSSM}

The discussion of EWSB must be expanded somewhat for the NMSSM due to the additional singlet degree of freedom in the Higgs sector; the introduction of a light singlet changes the vacuum structure of the potential and introduces a number of new parameters into the conditions for electroweak symmetry breaking. Fortunately, it is possible to develop a parametric understanding of the NMSSM vacuum in certain simplifying limits, and that will suffice for our purposes. We will find that the upshot remains largely the same as in the previous subsection -- for successful EWSB, we will need large negative $m_N^2$ at the weak scale, and there is a window of low messenger scales in which the negative, one-loop, $\Lambda/M$ suppressed contribution to $m_N^2$ ensures this. In what follows, our discussion will often mirror that of  \cite{Delgado:2007rz}, who considered this model without the large $A$-terms, and without the one-loop correction to $m_N^2$. We will see that these have vital effects and qualitatively change the behavior of the model.

 Upon introducing a singlet, the minimization conditions for the tree-level potential are extended to three equations: (\ref{eq:vacuummin1})-(\ref{eq:vacuummin2}), together with\footnote{Of course, these are merely the tree-level equations, which should be dressed with radiative corrections in the full solution. In what follows, we use \texttt{NMSSMTools} \cite{Ellwanger:2004xm, Ellwanger:2005dv} where necessary to capture the full effects of radiative corrections,  but find that the tree-level equations are adequate to understand the parametric behavior of the vacuum.}
 \begin{eqnarray}
\label{eq:vacuummin3}
2 \frac{\kappa^2}{\lambda^2} \mu^2 - \frac{\kappa}{\lambda} A_\kappa \mu + m_N^2 &=& \lambda^2 v^2 \left[ - 1 + \frac{1}{2} s_{2 \beta} \left( \frac{B_\mu}{\mu^2} + \frac{\kappa}{\lambda} \right) + \frac{1}{4} s_{2 \beta}^2 \frac{\lambda^2 v^2}{\mu^2} \right] ~.
\end{eqnarray}
These equations determine $\mu=\lambda\langle N\rangle$, and $B_\mu$ is given by:
\begin{equation}
\label{eq:Bmueq}
B_\mu = \frac{\kappa}{\lambda} \mu^2 - A_\lambda \mu - \frac{1}{2} s_{2 \beta} \lambda^2 v^2 ~.
\end{equation}

 In general, the solutions to (\ref{eq:vacuummin1})--(\ref{eq:vacuummin2}) and (\ref{eq:vacuummin3}) are complicated functions of the parameters and couplings.
However, things simplify considerably in the case of interest: $v^2 \ll m_{soft}^2$;  large $\tan \beta$ (to maximize the tree-level MSSM contributions to the Higgs mass); and $\lambda \ll 1$ (since we are not trying to lift the Higgs mass using the NMSSM quartic). This is a decoupling limit in which the singlet serves largely to fix $\mu$ and $B_\mu$ and does not mix significantly with the Higgs doublets. Consequently, the constraints on soft parameters imposed by EWSB in the MSSM are largely unchanged, but are supplemented by additional constraints on the singlet sector. In this regime, we find (provided $m_{H_d}^2 + \mu^2$ is not exceptionally large) the approximate relations
\begin{eqnarray}
\label{eq:mueqnmssm}  && \mu^2 \approx - m_{H_u}^2\\
   && B_\mu \approx  \mu \left( \frac{\kappa}{\lambda} \mu - A_\lambda \right) \approx 0\\
\label{eq:mnfroma}   && m_N^2 \approx A_\lambda (A_\kappa - 2 A_\lambda)
\end{eqnarray}
The third equation in particular is a parametrically interesting requirement: although $m_N^2$ may arise predominantly from either one-loop or two-loop contributions, depending on $\Lambda/M$, successful EWSB requires it be the same size as a two-loop contribution. But note that large $A_t$ automatically implies large $A_\lambda$ according to (\ref{alambda}), so $m_N^2$ at the weak scale must in general be quite large and negative for viable EWSB. The only possible exception is if there is some cancellation between $A_\kappa$ and $A_\lambda,$ in which case $m_N^2$ may be smaller and take either sign.

\begin{figure}[t]
\begin{center}
\includegraphics[width=3.5in]{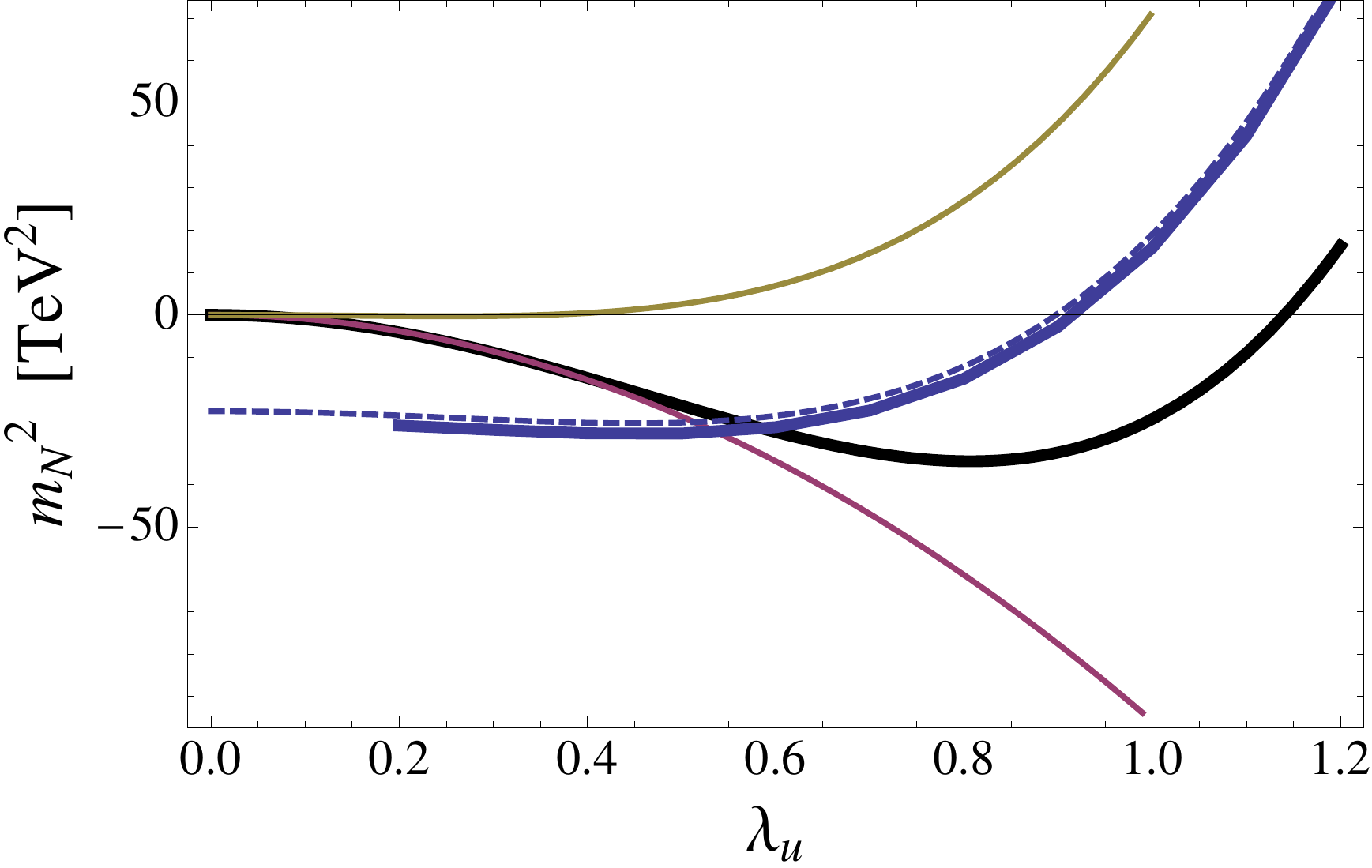}
\end{center}
\caption{Plot of $m_{N}^2$ vs.\ $\lambda_N$, for $\Lambda=110$ TeV, $M=220$ TeV, $\lambda_u=1.1$, $N_{mess}=4$, and $\kappa$ and $\lambda$ negligibly small. (At this point, the stops are about $2$ TeV and $m_h=125$ GeV.) The total value of $m^2_N$ at the messenger scale (\ref{eq:mNsq}) is shown in black; the red and yellow are the one and two loop contributions, respectively. Blue is the required value from EWSB. The dashed curve is given by the tree-level expression from (\ref{eq:mnfroma})  at the weak scale, and the solid curve is extracted from \texttt{NMSSMTools} at messenger scale. Clearly, both the RG and the weak-scale radiative corrections have negligible effects on $m_N^2$. The successful EWSB solution lies at where the blue and black curves intersect.  We see from this that the one-loop negative contribution to $m_N^2$ dominates and leads to a successful EWSB solution.   }
\label{fig:mNsqvslambdaN}
\end{figure}

What effects lead to large negative $m_N^2$ at the weak scale? Since we are focusing on low messenger scales and $\lambda\ll 1$, RG running is generally not sufficient. Instead, we are led to consider precisely the same mechanism as in the MSSM -- with moderate $\Lambda/M$, the negative, one-loop, $\Lambda/M$-suppressed contribution to $m_N^2$ in (\ref{eq:mNsq}) cannot be neglected, and can lead to a successful solution to the vacuum conditions. Note that the solution will generally prefer moderate $\lambda_N$ -- if $\lambda_N$ is too small, then $m_N^2$ is too small and cannot satisfy (\ref{eq:mnfroma}), while if $\lambda_N$ is too large, the (positive) two-loop term will dominate. The interplay of one- and two-loop contributions to $m_N^2$ and the requirements for a viable vacuum are illustrated in fig.\ \ref{fig:mNsqvslambdaN} for a particular point in parameter space.

\begin{figure}[t]
\begin{center}
\includegraphics[width=3in]{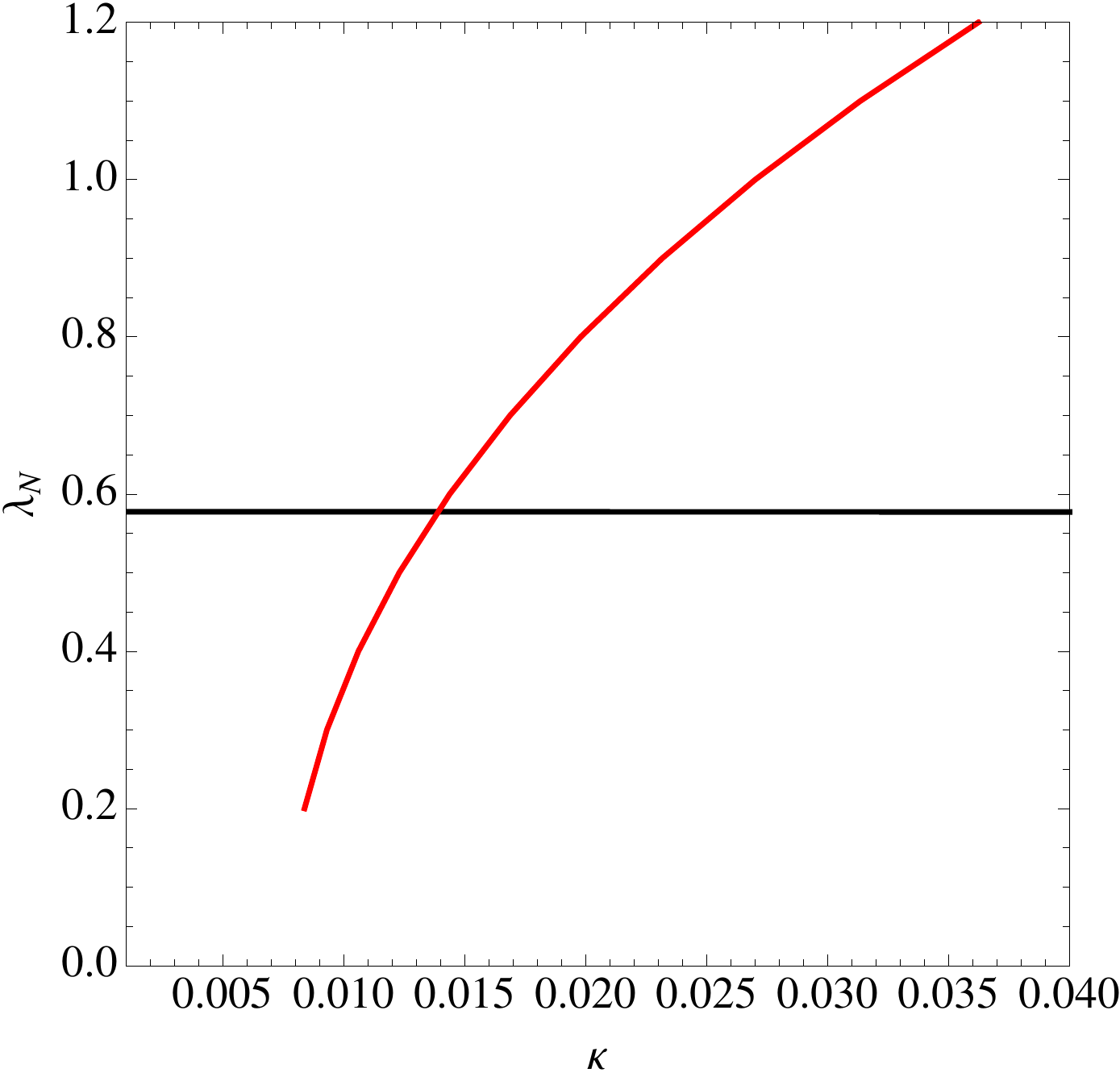}
\end{center}
\caption{Plot showing the EWSB solution in more detail, for the same parameter point as in fig.\ \ref{fig:mNsqvslambdaN}. The black line is the contour along which the input value of $m_N^2$ from the model (\ref{eq:mNsq}) agrees with the value required by EWSB (\ref{eq:mnfroma}). It is flat as a function of $\kappa$ because $\kappa$ has a negligible effect on $m_N^2$ when it is small. The red line indicates the agreement between input values and EWSB requirements for $\kappa$. The intersection of the two lines gives the consistent EWSB solution for a given value of $\Lambda, M, \lambda_u, N_{mess},$ and $\lambda$.}
\label{fig:nmssmintersection}
\end{figure}

As discussed in section \ref{sec:nmssm}, we can view the NMSSM model as adding three more parameters,  $(\lambda, \kappa, \lambda_N)$, to the MSSM module. The requirements for EWSB (\ref{eq:mueqnmssm})-(\ref{eq:mnfroma}) can be used to determine two of these parameters, say $(\kappa,\lambda_N)$, in terms of the third $(\lambda)$ and the other Higgs sector parameters. (In addition, the EWSB equations determine $\mu$.)
The nature of this solution is illustrated in fig.\ \ref{fig:nmssmintersection} for one point in parameter space, exemplifying our discussion of the parametric behavior of soft parameters required for EWSB. The red curve can be inferred to good accuracy by solving the approximate tree-level equations (\ref{eq:mueqnmssm})-(\ref{eq:mnfroma}) for $\kappa$ as a function of $\lambda_N$. The black curve meanwhile is flat as a function of $\kappa$ because when $\kappa$ is small it has essentially no effect on the dynamics of the model. In general, for fixed $\Lambda, M, \lambda_u, \lambda,$ and $N_{mess}$, each viable solution for EWSB constitutes a similar intersection of curves in the space of $\kappa$ and $\lambda_N$.

\subsection{Stop and slepton tachyons}

In addition to achieving successful EWSB, our models must also have a viable superpartner spectrum. In particular, the squarks and sleptons cannot be tachyonic at the weak scale. Weak-scale tachyons may  be induced either by direct contributions to the soft masses at the messenger scale  or by RG running.  These two effects provide further constraints on the parameter space.

 \begin{figure}[t]
 \begin{center}
\includegraphics[width=4.5in]{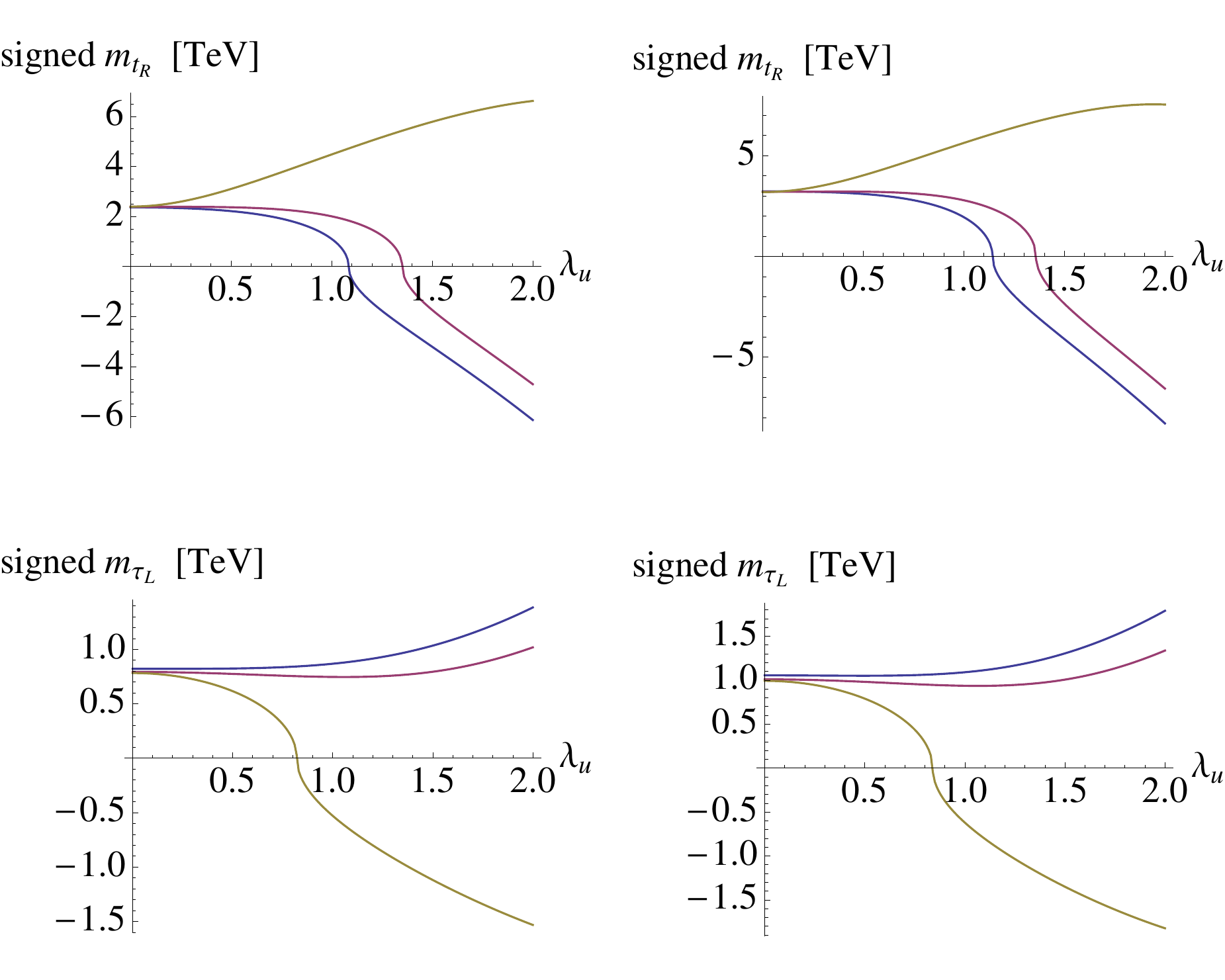}
\end{center}
\caption{Plot of the weak-scale signed masses $m_{\tilde t_R}^2/\sqrt{|m_{\tilde t_R}^2|}$ and $m_{\tilde \tau_L}^2/\sqrt{|m_{\tilde \tau_L}^2|}$ vs $\lambda_u$, for the ${\bf 5}\oplus {\bf\overline{5}}$ model (left) and the ${\bf 10}\oplus{\bf\overline{10}}$ model (right). The other parameter choices are as in fig.\ \ref{fig:mhuvslambdau}. We see that in both the ${\bf 5}\oplus{\bf\overline{5}}$ and  ${\bf 10}\oplus{\bf\overline{10}}$  models, tachyonic stops are problematic for $\lambda_u\gtrsim 1$, except for larger values of $\Lambda/M$. In this regime the stop mass gets pulled {\it up} during the RG flow by the large negative $m^2_{H_u}$. However, the same effect  pushes {\it down} the left-handed slepton masses, leading to tachyonic sleptons for  large $\Lambda/M$. Combined with fig.\ \ref{fig:mhuvslambdau}, we see that for $\lambda_u\sim 1$, a sweet spot exists with moderate $\Lambda/M$ where all constraints can  simultaneously be satisfied.}
\label{fig:msfermvslambdau}
\end{figure}

As we can see from (\ref{eq:mqsq}) and (\ref{eq:musq}), two-loop cross terms proportional to $\alpha_{\lambda_u}$  and $\alpha_t$ contribute negatively to the stop masses. Since $y_t \approx 1$ and $\lambda_u \gtrsim g_3$ in order to generate sufficiently large $A$-terms, this negative contribution is parametrically the same size (or larger!) than the positive gauge-mediated contribution.  Further, the large mixing will induce a bigger splitting between two mass eigenvalues in stop mass matrix.  These two effects lower the stop masses relative to those of other colored scalars. Avoiding prohibitive color-breaking minima therefore leads to an upper bound on $\lambda_u$.\footnote{Were it not for the Higgs mass, which in these models typically require stops above a TeV, this would be an amusing mechanism for generating a natural SUSY spectrum in gauge mediation. Even here, it typically renders the stops several hundred GeV lighter than other colored scalars.}

A completely different effect may lead to tachyonic sleptons. As discussed in section \ref{secEWSB}, EWSB in models with large $A$-terms and low messenger scales typically entails a large and negative contribution to $m_{H_u}^2$, already at the messenger scale.
This can lead to $m_{L}^2<0$ at the weak scale, through the hypercharge trace contribution to one-loop RGEs:
\begin{eqnarray}
\label{eq:RGLa}
& &  2\pi \frac{d}{d t} m_{L_a}^2 =  \delta_{a3} \alpha_{\tau} (m_{L_3}^2+m_{E_3}^2+m_{H_d}^2+A_{\tau}^2)- {3\over5}\alpha_1 M_1^2 -3 \alpha_2 M_2^2-\frac{3}{10}\alpha_1 \xi
\end{eqnarray}
where
\begin{eqnarray}
\label{eq:RGXi}
 & & \xi = {\rm Tr}[m_Q^2 -2 m_U^2 +m_D^2-m_L^2+m_E^2]+m_{H_u}^2-m_{H_d}^2
\end{eqnarray}
Since the stau mass eigenvalues are often further separated by mixing and the $\alpha_\tau$ contribution to the RGEs, typically the stau is the first state to be driven tachyonic. In any event, this translates to a requirement that $m_{H_u}^2$ cannot be too large and negative at the messenger scale.

Shown in fig.\ \ref{fig:msfermvslambdau} are plots of the stop and slepton masses, illustrating the interplay of all of these effects.

\subsection{Implications of the constraints}

Let us now conclude this section by summarizing briefly its main points and discussing their implications. In the previous subsections we have demonstrated that the soft parameters depend most sensitively on $\lambda_u$ and $\Lambda/M$. To achieve a large enough $A$-term for $m_h=125$ GeV, we need $\lambda_u\sim g_3\sim 1$. Then as we vary $\Lambda/M$, a number of issues can arise:
\begin{itemize}

\item $m_{H_u}^2$ (and $m_N^2$) can remain positive at the weak scale even after RG flow, preventing electroweak symmetry breaking. This occurs for low values of $\Lambda/M$.

\item $m_{H_u}^2$ can be too large and negative already at the messenger scale, resulting in a tachyonic sleptons at the weak scale due to the RG running. This problem arises for large values of $\Lambda/M$.

\item The stops can be negative at the messenger scale due to the direct contribution from the messenger-Higgs interactions. This effect happens at low to moderate $\Lambda/M$ and large $\lambda_u$.

\end{itemize}

Clearly a way to address the first two problems is to choose moderate values of the ratio $\Lambda/M$.  Note that since $\Lambda$ is usually fixed to be $\sim$ 100 TeV for reasonable superpartner masses, this implies that there is a ``sweet spot'' of $M \sim$  (a few) $\times $ 100 TeV where the model is viable. In this sense low messenger scales are actually an output of the model.

The third problem is still present at moderate $\Lambda/M$, and it is the ultimate limiting factor on the size of $A_t$.  Here the way out is to increase the messenger number -- $A_t$ and $m_{\tilde t}^2$ are both proportional to $N_{mess}$, and the relevant quantity for the Higgs mass is $A_t^2/m_{\tilde t}^2$.
 Indeed, for a model with ${\bf 5}\oplus{\bf\overline{5}}$ messengers, we find that for $N_{mess}=1$, $m_{h}=125$ GeV requires extremely heavy stops. But already for $N_{mess}=2$, $m_h=125$ GeV is possible for stops as light as 2 TeV. The situation improves somewhat for larger $N_{mess}$, though the improvement is saturated as the increasing messenger number also raises the gluino mass, which in turn pulls up the stop mass through RG flow. For ${\bf 10}\oplus{\bf\overline{10}}$ messengers, the effective messenger number already starts at three, so this is not an issue in this case. In the following section, we will focus on $N_{mess}=4$ for the ${\bf 5}\oplus{\bf\overline{5}}$ model and $N_{mess}=2$ for the ${\bf 10}\oplus{\bf\overline{10}}$ model.

\section{Spectrum and Phenomenology}
\label{sec:results}

In the previous section we discussed qualitatively the challenges and  possible solutions for viable models with large $A$-terms. In this section we will complete our analysis of these models by mapping out the available parameter space and phenomenology of  the ${\bf 5}\oplus {\bf\bar 5}$ and ${\bf 10}\oplus {\bf\overline{ 10}}$ benchmark models introduced in section \ref{sec:explicitmodel}.
In each case, we may consider either the simple MSSM module for $A$-terms, or the complete NMSSM theory that also generates $\mu$ and $B_\mu$. As discussed in section \ref{sec:EWSBNMSSM}, the vacuum structure of the NMSSM is more intricate, which will result in an additional constraint on the parameter space from requiring a nonzero singlet vev. Aside from this additional constraint, the analysis of the MSSM carries over completely to the NMSSM, since we are working in the decoupling limit where $\kappa\rightarrow0$ and $\lambda\rightarrow0$.  For numerical exploration of the parameter space, we use a combination of \texttt{softsusy v.3.3.0} \cite{Allanach:2001kg} and \texttt{NMSSMTools v.3.1.0} \cite{Ellwanger:2004xm, Ellwanger:2005dv}.

\subsection{Models with ${\bf 5 + \bar 5}$ messengers}
\label{sec:specificmodelfive}

\begin{figure}[t]
\begin{center}
\includegraphics[width=6in]{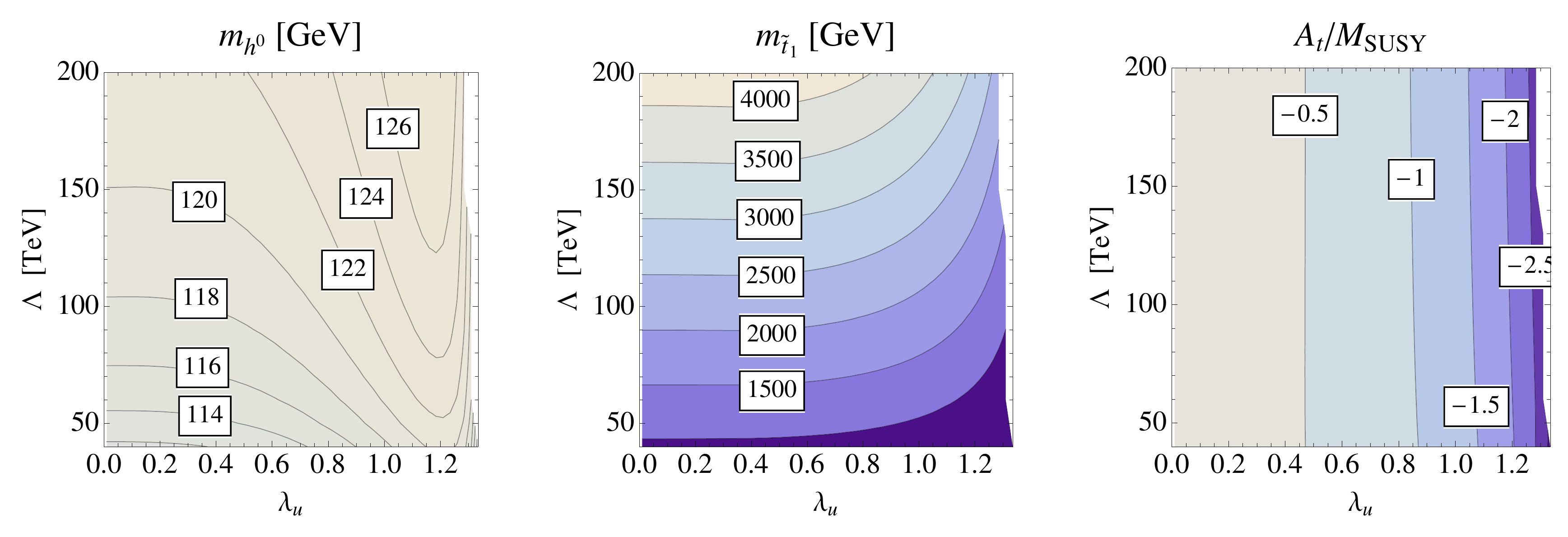}
\end{center}
\caption{Contour plots of $m_{h^0}$, $m_{\tilde t_1}$ and $A_t/M_{SUSY}$ in the $\Lambda$ vs.\ $\lambda_u$ plane, for $N_{mess}=4$ and $\Lambda/M=0.5$ (our best-case scenario for the ${\bf 5}\oplus{\bf\overline{5}}$ model).}
\label{fig:contourplots}
\end{figure}

As discussed in section \ref{subsec:mssmmod}, the parameter space of the MSSM version of this model consists of $\Lambda\equiv F/M$, $\Lambda/M$, $\lambda_u$, and $N_{mess}$. Since we are restricting ourselves to low-scale GMSB, we will only consider $N_{mess} \le 5$ to avoid Landau poles in the gauge couplings. In fig.\ \ref{fig:contourplots} we show contours of the Higgs mass as a function of \ $\Lambda$ and $\lambda_u$ in our ``best-case" model ($N_{mess} = 4, \Lambda / M  = 0.5$) in the ${\bf 5}\oplus {\bf\overline{5}}$ case; this choice of $N_{mess}$ and $\Lambda/M$ strikes a favorable balance between large $A$-terms and viable EWSB. We also show the variation in the lightest stop mass $m_{\tilde t_1}$ and the mixing ratio $A_t / M_{SUSY}$ (here $M_{SUSY} \equiv \sqrt{m_{\tilde t_1} m_{\tilde t_2}}$). The former is controlled mainly by $\Lambda$ (although as $\lambda_u$ increases, we see the approaching stop tachyon being reflected in the contours); while the latter is controlled by $\lambda_u$. We see that $m_h=125$ GeV is easily possible, with messenger scales as low as $\sim 100$ TeV and stops as light as 1500 GeV.

\begin{figure}[t]
\begin{center}
\includegraphics[width=6in]{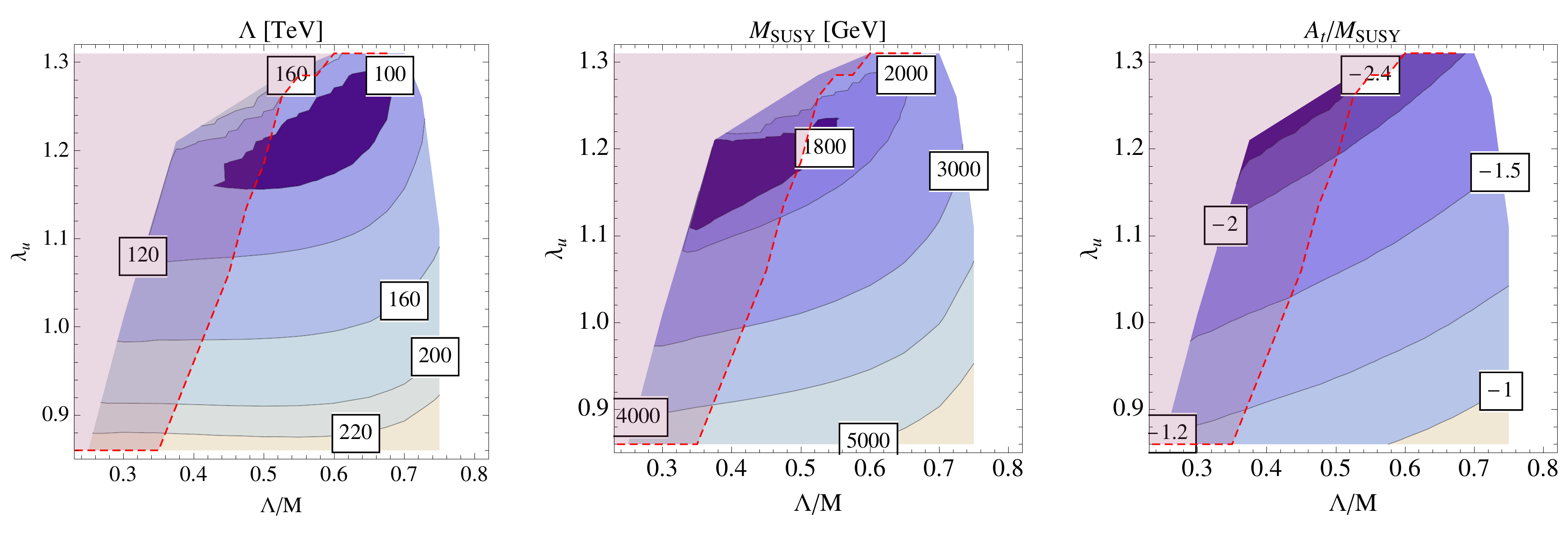}
\end{center}
\caption{Contour plots of the value of $\Lambda$ required for $m_h=125$ GeV in the $\lambda_u$ vs. $\Lambda/M$ plane for the ${\bf 5}\oplus{\bf\overline{5}}$ model, together with analogous plots for  $M_{SUSY}$ and $A_t/M_{SUSY}$. Here we have fixed $N_{mess}=4$. The white regions indicate where the spectrum runs afoul of tachyonic color/charge state or electroweak symmetry breaking. Overlaid in red is the region where there does not exist a consistent NMSSM solution with small $\lambda$.}
\label{fig:contourplots2}
\end{figure}

Since we are interested in $m_h=125$ GeV, it is perhaps more useful to focus on the subspace of parameters for which this is the case. Once we have fixed  the Higgs mass and chosen $N_{mess}=4$, the remaining parameter space of the model is precisely $(\Lambda/M,\lambda_u)$. So contour plots of quantities in this plane provide a complete characterization of the model for a given value of the Higgs mass. In fig.\ \ref{fig:contourplots2} we scan over the space of parameters for fixed $m_h = 125$ GeV as a function of $\Lambda/M$ and $\lambda_u$, showing contours of $\Lambda$, $M_{SUSY}$, and $A_t / M_{SUSY}$.  The viable parameter space is bounded by regions with  tachyonic superpartner masses or unsuccessful electroweak symmetry breaking, exemplifying our discussion in the previous section.

Generalization from the MSSM module to the full NMSSM model is straightforward.  As discussed in sections \ref{sec:nmssm} and \ref{sec:EWSBNMSSM}, the NMSSM introduces three new parameters ($\lambda, \kappa, \lambda_N$); we can choose to determine two of the extra parameters, say $\kappa$ and $\lambda_N$, in terms of the third ($\lambda$) and the other Higgs sector parameters.  So we only need to add one parameter, $\lambda$, to the MSSM parameter space.  Since our philosophy is to get $m_h=125$ GeV from stop mixing in the MSSM and  $\mu/B_\mu$ from the NMSSM, it is most favorable to operate in the decoupling limit $\lambda\ll1$; in the plots below we will take $\lambda=0.01$ for simplicity.

In section \ref{sec:EWSBNMSSM}, we also showed that viable EWSB in the NMSSM in the presence of large $A_t$ imposes the additional constraint that $m_N^2$ should be large and negative at the messenger scale to obtain a satisfactory $\mu$-term. For given values of $\Lambda/M$ and $\lambda_u$, there may not exist a value of $\lambda_N$ satisfying this constraint, in which case there is no consistent NMSSM solution. For instance, if $\Lambda/M$ is too small, $m_{H_u}^2>0$ as in the MSSM and/or we are unable to find a consistent NMSSM solution; typically the latter condition dominates. Meanwhile at high $\Lambda/M$, the stau is driven tachyonic as in the MSSM case. These constraints bound $0.35 \lesssim \Lambda / M \lesssim 0.8.$, and they remove a sizable chunk of the parameter space that is viable for the MSSM module alone.  In fig.\ \ref{fig:contourplots2} we have overlaid  the region in which there is no consistent NMSSM solution at small $\lambda$ onto the MSSM parameter space. The shape of this boundary in the plane of $\lambda_u$ and $\Lambda /M$ is approximately linear due to the conditions for obtaining a sufficiently negative value of $m_N^2$ in terms of $\lambda_u, \lambda_N,$ and $\Lambda/M$. While this certainly erodes some of the parameter space viable in the MSSM module,  a wide range of possible solutions remains.

\begin{figure}
\begin{center}
\includegraphics[width=6in]{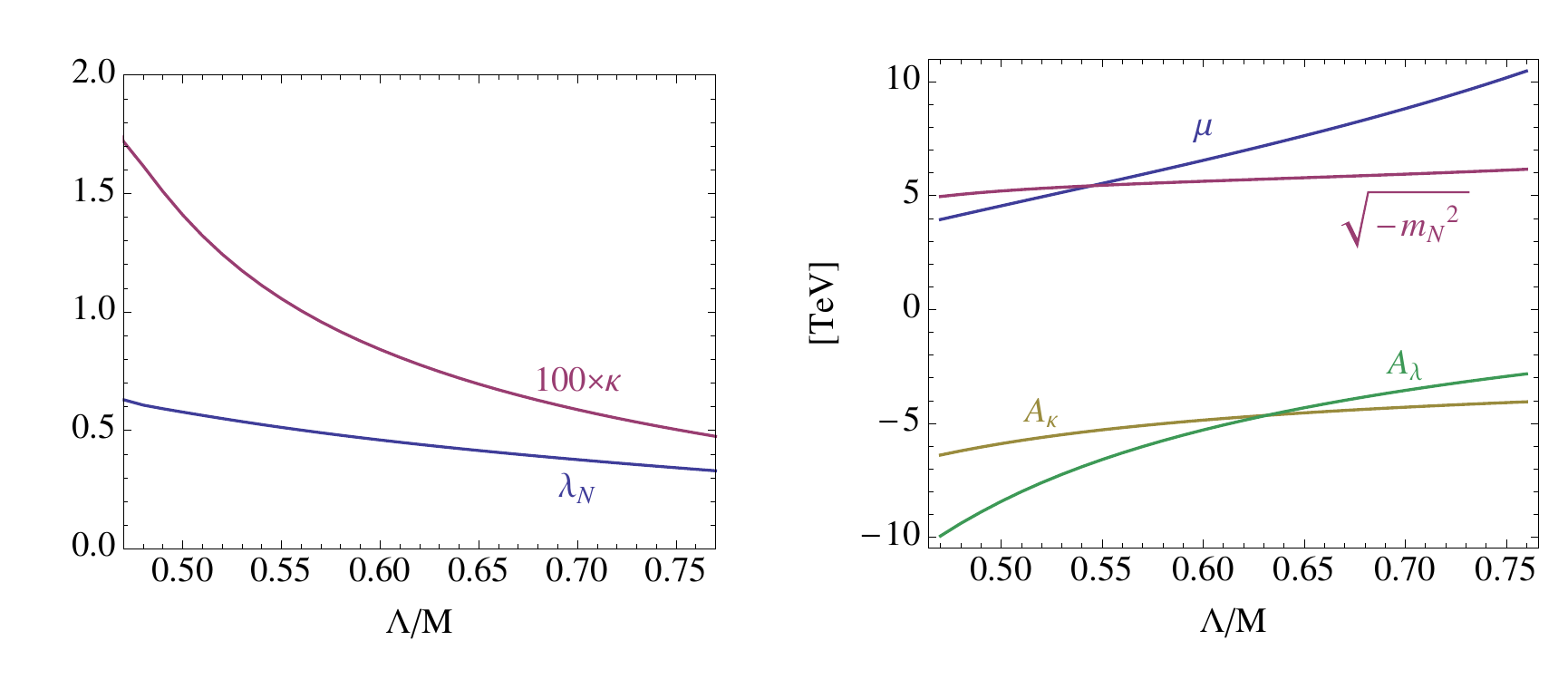}
\end{center}
\caption{A plot showing the dependence of the EWSB solution on $\Lambda/M$. The other parameters are as in fig.\ \ref{fig:mNsqvslambdaN}. Smaller values of $\Lambda/M$ are disallowed by $m_{H_u}^2>0$ at the weak scale and/or the inability to solve the $m_N^2$ equation for $\lambda_N$; while higher values are disallowed by $m_{\tilde\tau}^2<0$ at the weak scale.}
\label{fig:lambdaNkappavsLambdaNrat}
\end{figure}

Fig.\ \ref{fig:lambdaNkappavsLambdaNrat} exemplifies the discussion of section \ref{sec:challenges}, showing the EWSB solution for $\kappa$ and $\lambda_N$ as a function of the ratio $\Lambda/M$. The figure also shows the values of $\mu$ (determined by the vev of $N$) as a function of \ $\Lambda/M$. We see that as the ratio $\Lambda/M$ increases, $\mu$ likewise increases, because the large negative one-loop contribution to $m_{H_u}^2$ is taking over and must be cancelled by larger values of $\mu^2$ in order to yield the correct value of the Higgs vev.

Taken together, we see that a consistent model with $m_h=125$ GeV, a calculable source of $\mu$ and $B_\mu$, and a viable superpartner spectrum exists in a window around $\Lambda/M\sim 0.5$ where one-loop soft masses are important but not unreasonably large.

\subsection{Models with ${\bf 10 + \overline{10}}$ messengers}
\label{sec:specificmodelten}

For this type of model, the effective messenger number is automatically at least 3, which helps to increase stop mixing.  However, the effective messenger number increases rapidly with additional pairs of ${\bf 10}\oplus {\bf\overline{ 10}}$ messengers;  already with two pairs of ${\bf 10}\oplus {\bf\overline{10}}$, we are living dangerously at $M\sim 10^5$ GeV with regard to Landau poles for the Standard Model gauge couplings, as we discuss in more detail in appendix \ref{app:poles}. In general, enforcing perturbativity up to the GUT scale favors somewhat larger values of $M$ and thus smaller values of $\Lambda/M$.

\begin{figure}[t]
\begin{center}
\includegraphics[width=6in]{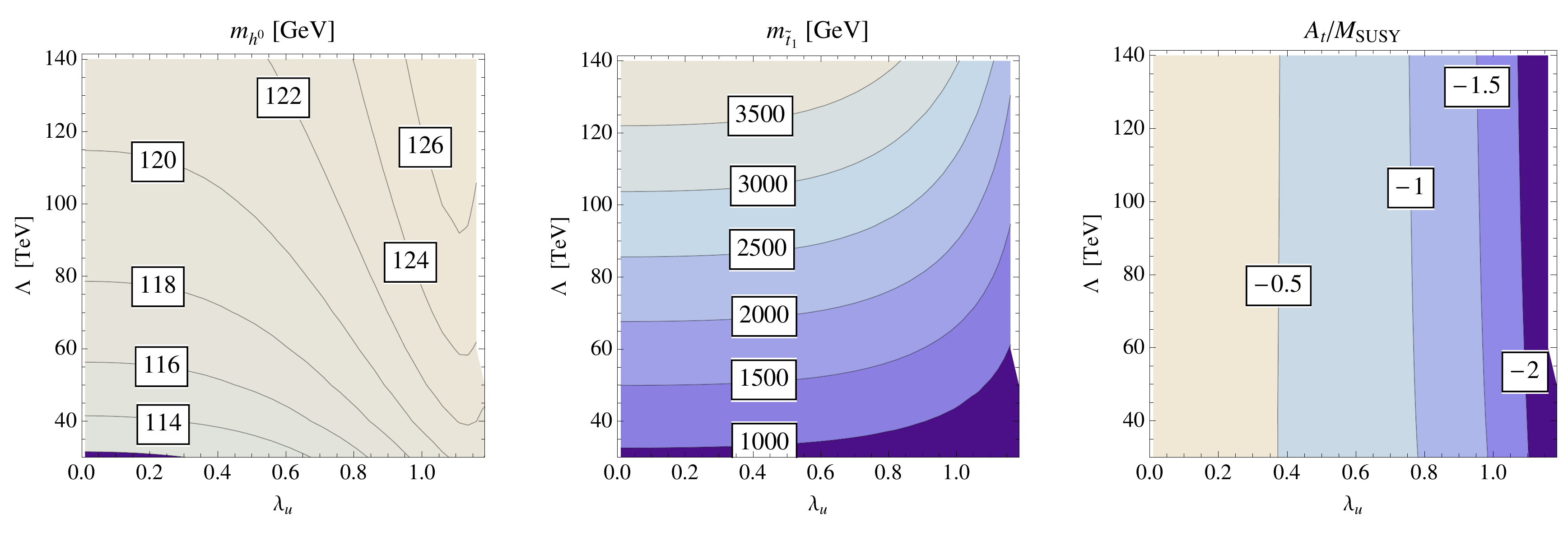}
\end{center}
\caption{Contour plots of $m_{h^0}$, $m_{\tilde t_1}$ and $A_t/M_{SUSY}$ in the $\Lambda$ vs.\ $\lambda_u$ plane, for $N_{mess}=2$ and $\Lambda/M=0.3$ (our best-case scenario for the ${\bf 10}\oplus{\bf\overline{10}}$ model).}
\label{fig:contourplots10}
\end{figure}

\begin{figure}[t]
\begin{center}
\includegraphics[width=6in]{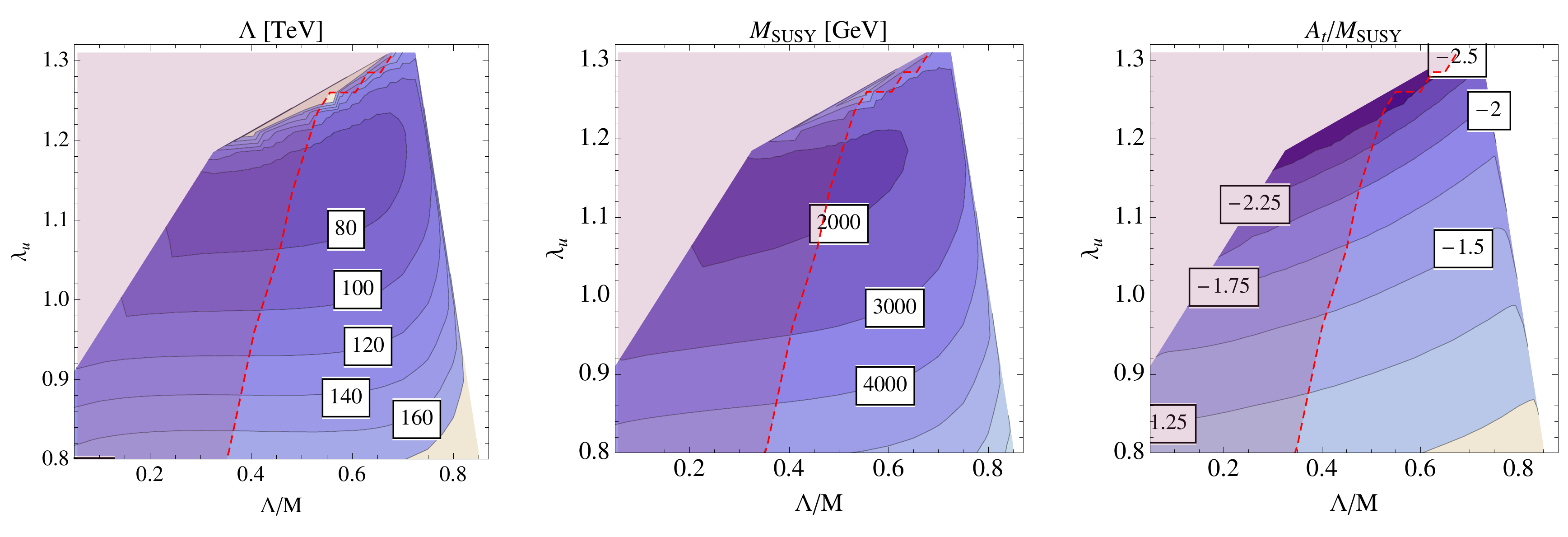}
\end{center}
\caption{Contour plots of the value of $\Lambda$ required for $m_h=125$ GeV in the $\lambda_u$ vs. $\Lambda/M$ plane for the ${\bf 10}\oplus{\bf\overline{10}}$ model, together with analogous plots for  $M_{SUSY}$ and $A_t/M_{SUSY}$. Here we have fixed $N_{mess}=2$. Overlaid in red is the region where there does not exist a consistent NMSSM solution with small $\lambda$. }
\label{fig:contourplots210}
\end{figure}

In fig.\ \ref{fig:contourplots10} we show contours of the Higgs mass in the plane of $\Lambda$ and $\lambda_u$. The contours of Higgs mass, stop mass, and the ratio $A_t / M_{SUSY}$ are qualitatively similar to the ${\bf 5} \oplus {\bf \bar 5}$ case. Fig.\ \ref{fig:contourplots210} shows the analogous contour plot of $\Lambda$ values required for $m_h = 125$ GeV in the plane of $\lambda_u$ and $\Lambda/M$. As we discussed in section \ref{secEWSB}, the region of viable solutions extends to much smaller values of $\Lambda/M$ than were allowed for  ${\bf 5} \oplus {\bf \bar 5}$ messengers, since the two-loop correction to $m_{H_u}^2$ is smaller and so the negative one-loop contribution is less important for successful EWSB. Indeed, at small $\Lambda/M$ there exists a sizable region where the Higgs soft masses are positive at the messenger scale and electroweak symmetry breaking occurs radiatively as in MGM.

As before, we may seamlessly generalize the MSSM module for ${\bf 10} \oplus {\bf\overline{10}}$ messengers to the full NMSSM.
The constraint imposed on $\Lambda / M$ by a viable solution for the NMSSM vacuum is parametrically similar to that in the case of ${\bf 5} \oplus {\bf\overline{5}}$ messengers in terms of the absolute limit, since the numerical details of the solution for NMSSM soft parameters are largely insensitive to the change in messenger representations. However, as shown in fig.\ \ref{fig:contourplots210}, the NMSSM vacuum constraint precludes the  region at small $\Lambda/M$ that is opened in the ${\bf 10} \oplus {\bf\overline{10}}$ case by reduced two-loop contributions to $m_{H_u}^2$.

\subsection{Phenomenology}
\label{sec:pheno}

\begin{figure}
\begin{center}
\includegraphics[width=4in]{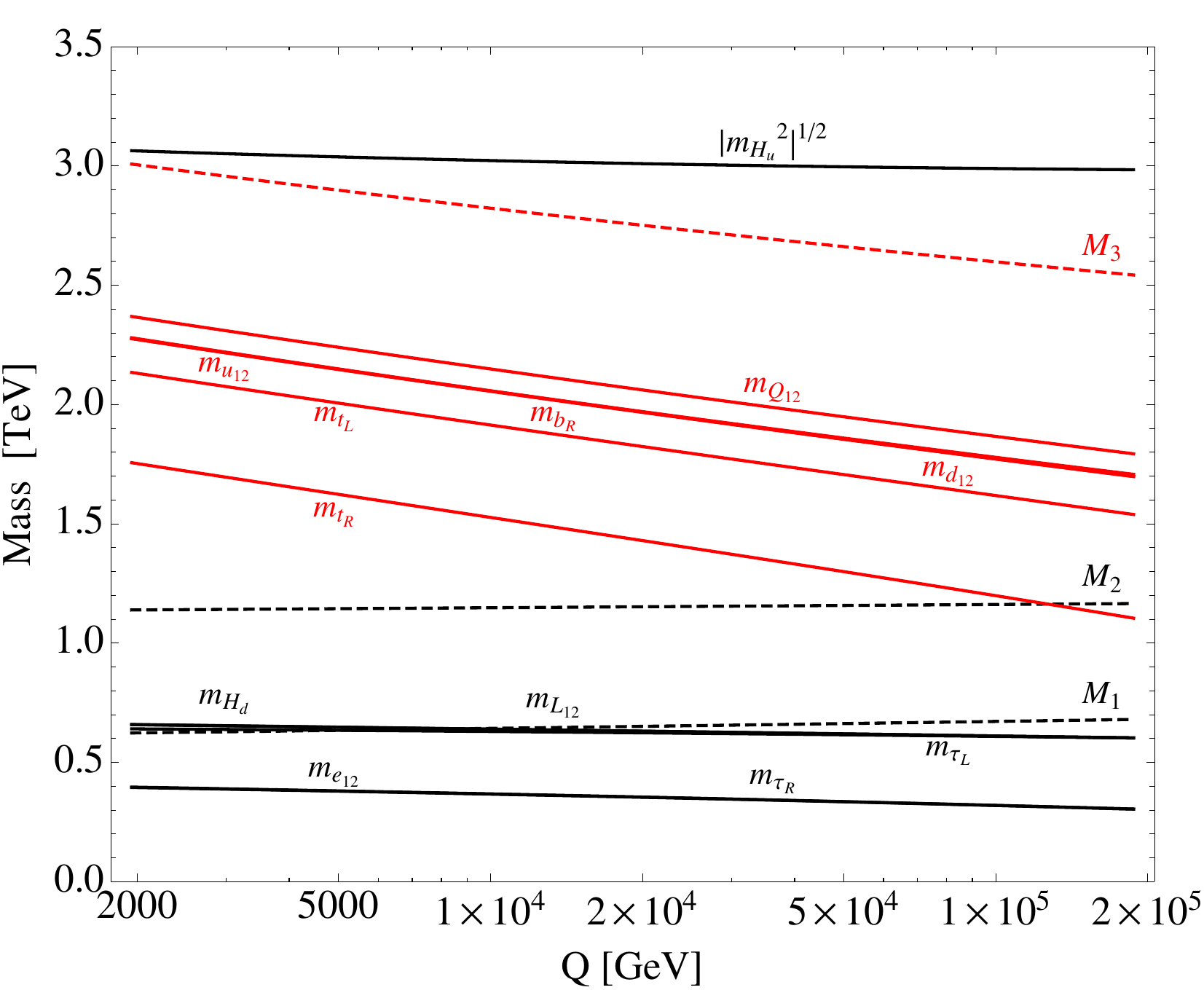}
\end{center}
\caption{Renormalization group evolution of a sample soft spectrum for $N_{mess} =2$, $\Lambda = 75$ TeV, $\Lambda / M = 0.4$, $\lambda_u = 1$ in the ${\bf 10}\oplus {\bf\overline{10}}$ case (for which $m_h=125$ GeV). Soft masses for colored superpartners are shown in red, while those for electroweak superpartners are shown in black. Dashed lines denote gaugino masses, solid lines denote scalar masses. Note the $H_u$ soft  mass-squared is negative. }
\label{fig:samplerg}
\end{figure}

\begin{figure}
\begin{center}
\includegraphics[width=5in]{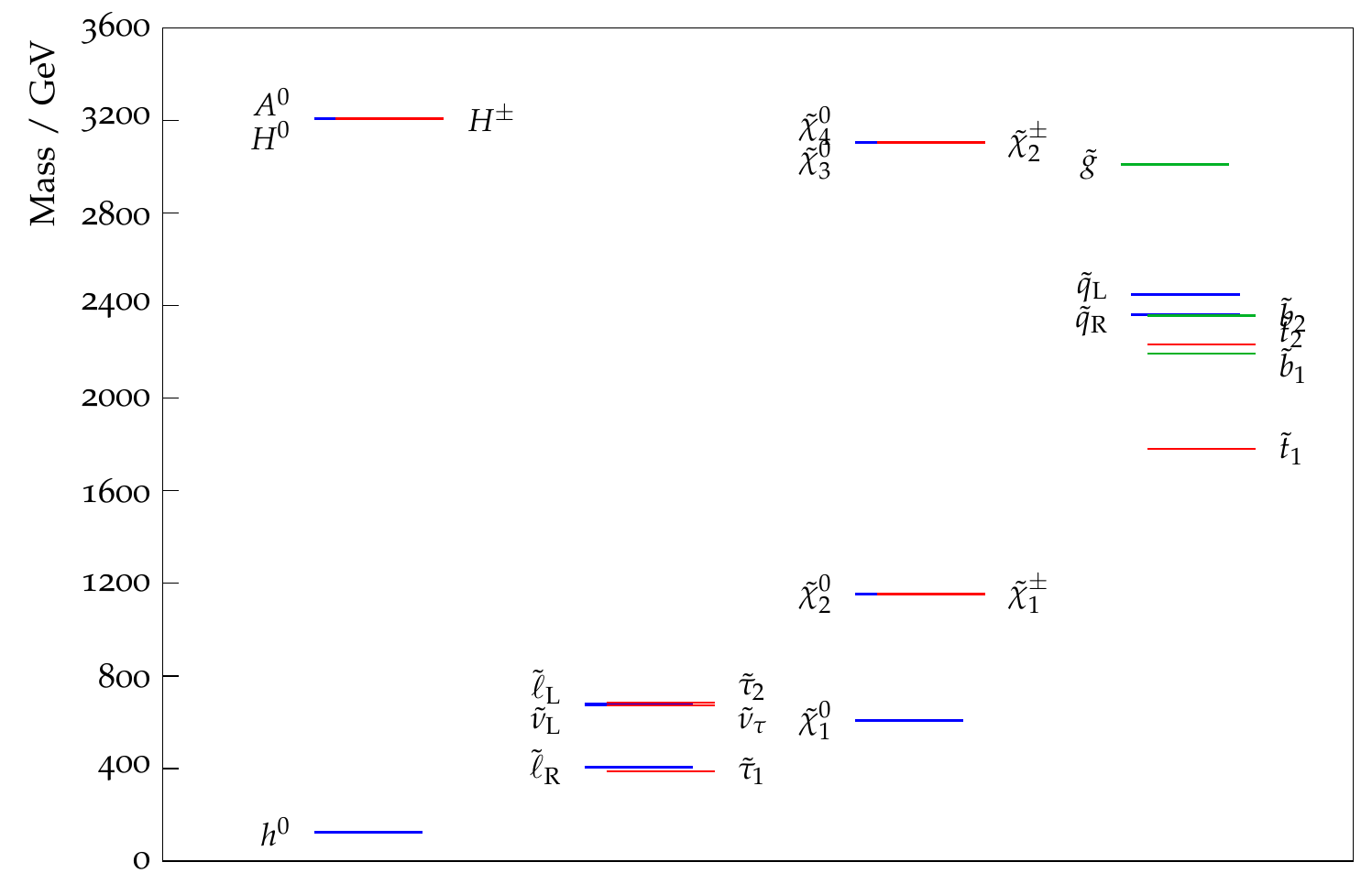}
\end{center}
\caption{The physical mass spectrum, for the same model as in fig.\ref{fig:samplerg}.
\label{55spectrumExample}}
\end{figure}

Finally, let us briefly describe the phenomenology of the models considered in the previous sections. The low-energy spectrum does not differ radically between the MSSM and NMSSM cases, since in the NMSSM models the singlet degrees of freedom are heavy and decoupled. Similarly, there are only slight differences between the choice of messenger representations, up to the general effects of changing the effective messenger number. The RG evolution of a  representative  soft spectrum is shown in fig.\ \ref{fig:samplerg}, while the mass spectrum for this point is shown in fig.\ \ref{55spectrumExample}.

In the colored sector, the stops are typically lighter than in minimal GMSB for the same value of $\Lambda$ and $N_{mess}$, due to the additional negative contributions from Higgs-messenger couplings. The large $A$-term also increases the splitting between the two stop mass eigenvalues, which further lowers the mass of the lighter eigenstate $\tilde{t}_1$. Even so, attaining $m_h = 125$ GeV typically requires stops above $\sim 1.5$ TeV and gluinos above 2 TeV (and more typically near 3 TeV). Thus the cross section for colored sparticle production is typically quite low, near the limit of observability at the LHC.

In the electroweak sector, the sleptons and electroweakinos are typically at or below a TeV, with the usual MGM splitting between the wino and bino. The sleptons are typically lighter than the wino. The cross section for electroweak sparticle production is also quite low, but nonetheless observable at the LHC. Note that the higgsinos and Higgs scalars $H^0, A^0,$ and $H^\pm$ are quite heavy due to the large value of $\mu$ necessitated by EWSB, so that the Higgs sector is far into the decoupling limit and the lightest Higgs properties are those of the Standard Model.

The NLSP is almost invariably the stau, except in very small regions of parameter space where it may become a mostly-bino neutralino. The staus are heavily mixed, such that the lightest stau is always lighter than the sneutrino $\tilde \nu_\tau$ and there is no co-NLSP. Since the scale of SUSY breaking is low in these scenarios, the NLSP decays promptly in the detector; the most promising search channels for this spectrum are likely to be those involving leptons plus missing transverse energy, such as the $H_T/E_T^{miss}$ binning of the  CMS multilepton search \cite{Chatrchyan:2012ye}.  In that paper, limits were set on a GMSB-motivated benchmark model  which has degenerate slepton co-NLSPs and specific mass relations among the superpartners.  So as such, it is not possible to directly use the CMS search to infer limits on our scenario, which has stau NLSP. It would be interesting to recast the CMS search in terms of our model; this should be straightforward, since they provide the data for channels where taus are included. Furthermore, we expect that the limits are strictly weaker for stau NLSPs compared to slepton co-NLSPs. For decoupled squarks and gluinos, the CMS limit was $m_{\tilde \chi_1^\pm}\gtrsim 600$ GeV, with $m_{\tilde\ell_R}=0.3m_{\tilde\chi_1^\pm}$, $m_{\tilde\chi_1^-} = 0.5 m_{\tilde\chi_1^\pm}$, and $m_{\tilde\ell_L}=0.8m_{\tilde\chi_1^\pm}$. So we are confident that the existing search does not yet meaningfully encroach on our parameter space. Nevertheless, multilepton searches should ultimately prove sensitive with increased integrated luminosity.

Although ancillary to the phenomenology, we conclude with a few
remarks on fine-tuning in the EWSB potential given this
characteristic spectrum. As usual, at large $\tan \beta$ the tuning
in the potential is governed by a cancellation between $\mu^2$ and
$m_{H_u}^2$. In both the MSSM and NMSSM models, the
overall tuning (as quantified by the Barbieri-Giudice measure
\cite{Barbieri:1987fn}) is typically $10^3 - 10^4$. The
tuning in the NMSSM is generically larger than that in the MSSM since the NMSSM has
a stronger constraint from EWSB that excludes some of the less tuned points.  
In section \ref{subsec:littleamh} we introduced the little $A$ - $m_H^2$ problem,
which is essentially the observation that $m_{H_u}^2$ always
receives an irreducible, positive contribution from $A_u^2$. This large contribution
must either cancel against a large $\mu^2$-term or against another
large term in \ref{eq:mhusqfinal} with opposite sign. Either way,
such a large cancellation greatly enhances the tuning of the model.

In the MSSM models, the tuning is therefore dominated by $\lambda_u$, which controls both $A_t$ and the one- and two-loop contributions to $m_{H_u}^2$.  In the NMSSM, the situation is similar, though now the tuning associated with $\mu$ is translated to a tuning in $\lambda_N, \lambda$, and $\kappa$ via (\ref{eq:vacuummin3})-(\ref{eq:Bmueq}). Amusingly, there is very little tuning associated with the scale of colored superpartners since the threshold contributions to Higgs soft parameters from Higgs-messenger couplings are far more important than radiative corrections from colored scalars. In this sense, the relative heaviness of the gluino and squarks is a red herring for tuning at the weak scale in these models.

\section{Conclusions}
\label{sec:conclusions}

A Higgs boson at 125 GeV poses a challenge for the MSSM in general and in gauge mediation in particular. If colored superpartners lie within reach of the LHC, explaining the Higgs mass requires large $A$-terms that are unavailable to pure gauge mediation unless the messenger scale is high \cite{Draper:2011aa}. This constraint would appear to challenge the possibility that supersymmetry may be broken and mediated to the MSSM at relatively low scales. Yet low-scale gauge mediation remains an attractive framework due to its distinctive phenomenology, including features such as favorable gravitino cosmology and prompt NLSP decays. This strongly motivates exploring ways in which low-scale gauge mediation might be reconciled with the presumptive mass of the Higgs.

In this work we have constructed simple, economical, and calculable models of low-scale gauge mediation that generate all the necessary parameters in the Higgs sector of the MSSM and readily provide a Higgs at 125 GeV. The key feature is the introduction of Higgs-messenger interactions that lead to large $A$-terms aligned with the SM flavor structure. This is a natural step in the context of GMSB, since the $\mu$ - $B_\mu$ problem already suggests that additional interactions are required in the Higgs sector. In general, such interactions lead to an $A$ - $m_H^2$ problem, which is solved if the only source of mass in the messenger sector is the expectation value of a single SUSY-breaking spurion (i.e.\ if the messengers are described by minimal gauge mediation). Such models suffice for generating large $A$-terms and stop mass mixing required for the Higgs, but on their own do not solve the $\mu$ - $B_\mu$ problem. In this sense they constitute ``modules'' that may be appended to other solutions to the $\mu$ - $B_\mu$ problem. One particularly compelling solution is in the context of the NMSSM, where a simple generalization to include singlet-messenger couplings simultaneously ameliorates the problems of the NMSSM in GMSB, and generates viable $\mu$ and $B_\mu$. Since the Higgs mass arises primarily due to stop mixing, the singlet sector serves only to generate $\mu$ and $B_\mu$, thereby avoiding problematically large singlet-Higgs couplings with Landau poles at low scales. Indeed, these theories remain weakly-coupled up to, and generally well beyond, the messenger scale. It is compelling that a straightforward generalization of low-scale gauge mediation to include perturbative interactions between the Higgs sector and messenger sector -- interactions already hinted at by the $\mu$ - $B_\mu$ problem -- naturally accommodates a Higgs at 125 GeV and provides all necessary soft parameters.


Our approach builds on previous works, especially \cite{Delgado:2007rz} and  \cite{Kang:2012ra}, but the complete combination of interactions for large $A$-terms and $\mu/B_\mu$, and the emphasis on low messenger scales, are both novel and lead to a qualitatively new model with distinctive features.  Chief among these is the crucial role played by one-loop $\Lambda/M$-suppressed contributions to $m_{H_u}^2$  in guaranteeing electroweak symmetry breaking, which is otherwise imperiled by large two-loop soft masses that accompany sizable $A$-terms.

The phenomenology of these models is very similar to that of MGM with high effective messenger number. One notable difference is that the mass of the stop is always significantly lowered relative to the masses of other colored scalars due to the Higgs-messenger interactions, such that the lightest stop is typically several hundred GeV lighter than the remaining squarks. Even so, a Higgs mass at $m_h = 125$ GeV suggests the stop is relatively heavy on LHC scales, above $\sim 1.5$ TeV, with the gluino above 2 TeV. The NLSP is almost always the stau, though in some cases it may be the lightest neutralino. In either case, NLSP decays to the gravitino are always prompt due to the low messenger scale. Overall, the spectrum is quite consistent with current collider limits and perhaps explains why we have yet to observe evidence for supersymmetry at the LHC.

There are numerous avenues for future study. The models presented here have a potentially large parameter space, of which we have only considered a simplified subspace. It would be interesting, for example, to study the consequences of splitting the messenger multiplets on the low-energy phenomenology. More generally, it should be possible to construct weakly-coupled models with large $A$-terms that realize the full parameter space allowed by general gauge mediation \cite{Meade:2008wd,Buican:2008ws}, which would allow for a greater range of NLSP candidates and collider signals. We have also focused exclusively on the decoupling limit of the NMSSM, where $\lambda, \kappa \ll 1$; it may be the case that other parametric regimes are allowed, in which case Higgs signals could deviate from Standard Model expectations. Finally, we have remained agnostic to the origin of the supersymmetry breaking and messenger sectors. Ultimately, it is worth exploring whether our models might  be embedded in a complete theory of dynamical supersymmetry breaking in which Higgs-messenger couplings are a natural ingredient.

It bears emphasizing we have limited our focus to weakly-coupled theories with perturbative messenger sectors and decoupled hidden sector interactions. It is plausible that the related problems of $\mu$ -$B_\mu$, $A$ - $m_H^2$, and $m_h = 125$ GeV may alternatively be resolved in a strongly-coupled hidden sector along the lines of \cite{Dine:2004dv,Murayama:2007ge, Roy:2007nz}. In this case, the details of the hidden sector interactions are crucial to the boundary conditions for soft parameters \cite{Craig:2009rk}, and it would be interesting to systematically study implications for the Higgs sector in terms of hidden- and messenger-sector correlation functions \cite{ourwork}.

\section*{Acknowledgments}
We thank  U.~Ellwanger, S.~Thomas, and F.~Zwirner for useful conversations. NC is supported by NSF grant PHY-0907744, DOE grant DE-FG02-96ER40959, and the Institute for Advanced Study. SK and YZ are supported by DOE grant DE-FG02-96ER40959. DS is supported in part by a DOE Early Career Award and a Sloan Foundation Fellowship.

\appendix

\section{General Formulas}
\label{app:genmass}

Whenever SUSY breaking may be parameterized by a single spurion $X$ whose lowest expectation value is responsible for messenger masses, the soft spectrum may be computed to leading order in $\Lambda/M$ via analytic continuation into superspace \cite{Giudice:1997ni}.  The resulting soft masses and $A$-terms for arbitrary marginal visible-messenger superpotential interactions linear in the visible sector fields are \cite{Chacko:2001km}:\footnote{The conventions for $A_a$ and the anomalous dimensions used in \cite{Chacko:2001km} differ slightly from ours. We are using the conventions of \cite{Martin:1997ns}.}
\begin{align}
\delta m_a^2\Big|_{t=\frac{1}{2}\log |M|^2} &= \frac{1}{2}\left[\sum_m\big( \frac{d\gamma_a^+}{d\alpha_m}- \frac{d\gamma_a^-}{d\alpha_m}\big)\partial_t \alpha^+_m- \frac{d\gamma_a^-}{d\alpha_m}\big(\partial_t \alpha^+_m-\partial_t \alpha^-_m\big)\right]\Lambda^2\Big|_{t=\frac{1}{2}\log |M|^2}\label{appMh}\\
A_a\Big|_{t=\frac{1}{2}\log |M|^2}&=-\sum_m\Big(\frac{d\gamma_a^+}{d\alpha_m}- \frac{d\gamma_a^-}{d\alpha_m}\Big) \alpha^+_m \Lambda\Big |_{t=\frac{1}{2}\log |M|^2}\label{appAt}
 \end{align}
where the $\delta$ denotes a correction to the usual GMSB soft masses. The $A$-term computed here correspond to a specific field label by $a$, rather than to a coupling. The $A$-terms corresponding to the couplings $y_t,\; \kappa,\; \lambda,\; \dots$ are linear combinations of the $A$-terms computed in (\ref{appAt}).  In what follows $i$, $j$, etc.\ range over messenger fields; and $a$, $b$, etc.\  range over visible sector  fields. Repeated indices are summed over, except for the free index $a$. The $A$-terms appear in the potential via $V \supset A_a \phi_a \partial_{\phi_a} W(\phi) + {\rm h.c.}$. The $\gamma_a^\pm\equiv-\frac{1}{2}\frac{\partial \log Z_a^\pm}{\partial t}$ and $\alpha^\pm_m\equiv\frac{(\lambda_m^\pm)^2}{4\pi}$ are the anomalous dimensions and couplings above and below the messenger threshold, respectively. The sum over $m$ runs over all the couplings in the theory.

We convert (\ref{appMh}) and (\ref{appAt}) into more explicit formulas by specifying the anomalous dimensions and the $\beta$ functions, accounting for couplings between messengers and matter fields but neglecting possible couplings between messengers alone.  The anomalous dimensions are then given by
\begin{eqnarray}
 &\gamma^a={1\over4\pi}\Big(\frac{1}{2}d^{ij}_a\alpha_{aij}+{1\over2}d^{bc}_a\alpha_{abc}-2 c^a_r \alpha_r \Big)\\
&  \gamma^i=\frac{1}{4\pi}\Big(d^{aj}_i\alpha_{aij}-2 c^i_r \alpha_r \Big)
 \end{eqnarray}
 The $d_{a}^{ij}$ count the number of fields $i$, $j$ talking to $a$ through the Yukawa vertex $(aij)$. With a slight abuse of notation, we denote the couplings with the messenger sector, the MSSM Yukawa couplings, and the gauge couplings with  $\alpha_{aij}$, $\alpha_{abc}$ and $\alpha_r$ respectively. Similarly, the relevant beta functions are
\begin{eqnarray}
 \partial_t \alpha^\pm_{abc}=&2 \alpha_{abc}(\gamma_a^\pm+\gamma_b^\pm+\gamma_c^\pm)\\
 \partial_t\alpha^\pm_{aij}=&2\alpha_{aij}(\gamma_a^\pm+\gamma_i^\pm+\gamma_j^\pm)
 \end{eqnarray}
 where the $\pm$ subscript again indicates whether $\alpha$ and $\gamma$ are to be taken above or below the messenger threshold. Substituting these formulas in (\ref{appMh}) and (\ref{appAt}) yields
\begin{eqnarray}\label{eq:msqgen} \nonumber
&& \delta m_a^2  = {1\over8\pi^2}\left(-{1\over2}d_a^{ij}(c_r^a+c_r^i+c_r^j)\alpha_r \alpha_{aij}+{1\over8}\left(d_a^{ij}\alpha_{aij}\right)^2+{1\over2}d_{a}^{ij}d_{i}^{bk}\alpha_{bik}\alpha_{aij}
 - {1\over4}d_{a}^{bc}d_{b}^{ij}\alpha_{abc}\alpha_{bij}\right)\Lambda^2 \\
&& A_a = -{1\over 8\pi}d_a^{ij}\alpha_{aij}\Lambda ~.\label{eq:msqgenAterm}
\end{eqnarray}

Now to obtain the formulas (\ref{eq:mhusqfinal})-(\ref{eq:Atfinal}) and (\ref{msqfinalnmssm})-(\ref{akappa}) in the bulk of the text, it suffices to substitute for the correct $d_a^{ij}$, $\alpha_{aij}$ and $\alpha_{abc}$. For the MSSM, the indices $a,b,c,\dots$ run over the fields $H_u$, $Q$ and $U$. The indices $i,j,k,\dots$ run over the messenger fields $\phi_i$ and $\tilde\phi_i$. With this in mind, one can read off the non-zero $d$'s and the couplings from (\ref{eq:gensuper}):
\begin{equation}
d_{H_u}^{\phi_1\tilde\phi_2}= d_H,\qquad  d_{\phi_1}^{H_u \tilde\phi_2}+d_{\tilde\phi_2}^{H_u \phi_1}=3,\qquad \alpha_{H_u\phi_1\tilde\phi_2}= \alpha_{\lambda_u},\qquad  \alpha_{H_u QU}= \alpha_t
 \end{equation}
 with $d_H$ given by (\ref{eq:dcfive}) or (\ref{dcten}) in the ${\bf 5}\oplus{\bf\overline{5}}$ or ${\bf 10}\oplus{\bf\overline{10}}$ models respectively.
The same conventions hold for the NMSSM, with the important difference that the indices $a,b,c,\dots$ now can take the value $H_d$ and $N$ as well. Moreover several extra couplings must be accounted for:\footnote{Note the extra factor of 4 in the translation of $\alpha_{NNN}$ to $\alpha_\kappa$. This is because of the non-standard NMSSM convention for the normalization of $\kappa$.}
\begin{equation}
\begin{array}{lll}
 d_{H_u}^{\phi_1\tilde\phi_2}+d_{H_u}^{\varphi_1\tilde\varphi_2} = d_H,  &\quad d_{\phi_1}^{H_u \tilde\phi_2}=d_{\varphi_1}^{H_u \tilde\varphi_2}=d_1^{H2}, &\quad   d_{\tilde\phi_2}^{H_u \phi_1}=d_{\tilde\varphi_2}^{H_u \varphi_1}=d_2^{H1},\smallskip \\
  d_N^{\varphi_1\tilde\phi_1}+d_N^{\varphi_2\tilde\phi_2}  =d_N, &\quad d_{\phi_1}^{N\tilde\varphi_1}= d_{\phi_2}^{N\tilde\varphi_2}=d_{\tilde\varphi_1}^{N\phi_1}=d_{\tilde\varphi_2}^{N\phi_2}=1, & \smallskip\\
d_N^{H_uH_d}=2,  &\quad d_{H_u}^{NH_d}=1, &\quad d_N^{NN}=1, \smallskip\\
 \alpha_{H_u\phi_1\tilde\phi_2}= \alpha_{H_u\varphi_1\tilde\varphi_2} = \alpha_{\lambda_u}, &\quad \alpha_{N\phi_i\tilde\varphi_i} =\alpha_{\lambda_N} & \smallskip\\
  \alpha_{H_u QU}=\alpha_t ,&\quad  \alpha_{NH_uH_d}=\alpha_\lambda, &\quad \alpha_{NNN} = 4\alpha_\kappa
\end{array}
 \end{equation}
with $d_H$, $d_N$, etc.\ again given in  (\ref{eq:dcfive}) or (\ref{dcten}). Finally note that equation (\ref{eq:msqgenAterm}) computes the $A$-terms corresponding to various fields (see (\ref{eq:Higgspars})), instead of couplings. The $A$-term for the various couplings that are used in the bulk of the draft can be obtained as follows:
 \begin{eqnarray}
 A_t &=& A_{H_u} \nonumber\\
 A_\lambda &=& A_N+A_{H_u} \nonumber\\
 A_\kappa &=&3A_N.
 \end{eqnarray}

\section{Physics Above the Messenger Scale}
\label{app:poles}

The models presented in sections \ref{sec:specificmodelfive} and  \ref{sec:specificmodelten} are complete and calculable effective theories below the messenger scale. This is the most that one can concretely ask for when treating the hidden sector in the spurion limit, since above the messenger scale the dynamics of the hidden sector and the origin of hidden-messenger couplings are bound to become important. However, one may still wish to study the behavior of the theory above the messenger scale, modulo ignorance of hidden sector dynamics.

Unlike in many realizations of the NMSSM where the singlet contributions to the potential are used to raise the Higgs mass, there is no problem with Landau poles in $\lambda$ for any of the models we consider. Such Landau poles would be particularly troublesome since they involve all the light degrees of freedom in the EWSB sector. Since we are working in the decoupling limit, $\lambda$ is always very small at the weak scale, and although it grows in the ultraviolet, it easily remains perturbative all the way to the GUT scale. The same may be said of $\kappa$, which is likewise small at the weak scale and never runs large below the GUT scale. Thus all the parameters in the NMSSM effective theory below the messenger scale are well-behaved above it as well.

On the other hand, there may conceivably be Landau poles in the gauge couplings and the couplings introduced at the messenger scale. The particular complications are qualitatively different depending on the messenger representations. For MSSM and NMSSM models with ${\bf 5 \oplus \overline{5}}$ messengers there are no irreducible Landau poles in the Standard Model gauge couplings up to the GUT scale for any value of the messenger scale, since viable models exist with $N_{mess} \leq 4$. For models with ${\bf 10 \oplus \overline{10}}$ messengers there may be Landau poles in the Standard Model gauge couplings before the GUT scale due to the large effective messenger number if the messenger scale is too low. However, for the most minimal NMSSM model (with effective messenger number 6) we find there are no Landau poles across the range of messenger scales under consideration, as determined by two-loop RG running and one-loop threshold matching. But the Standard Model gauge couplings grow strong as they approach the GUT scale, and perturbation theory is perhaps no longer reliable. Higher values of the effective messenger number, corresponding to more than two pairs of ${\bf 10 \oplus \overline{10}}$, introduce Landau poles in the gauge couplings below the GUT scale.

\begin{figure}[t]
\begin{center}
\includegraphics[width=2in]{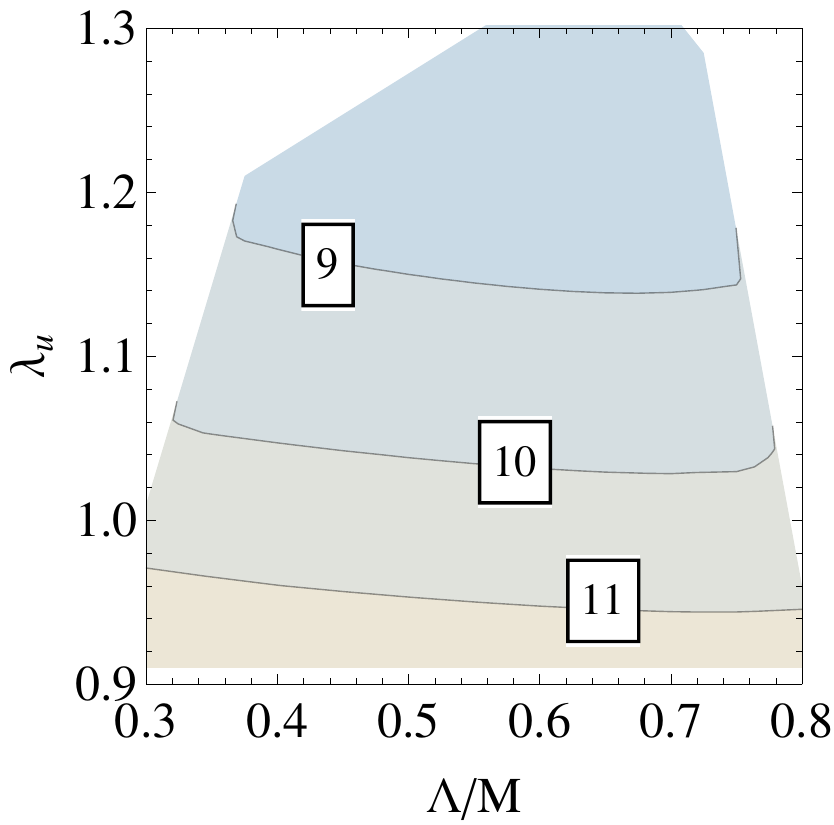}
\includegraphics[width=2in]{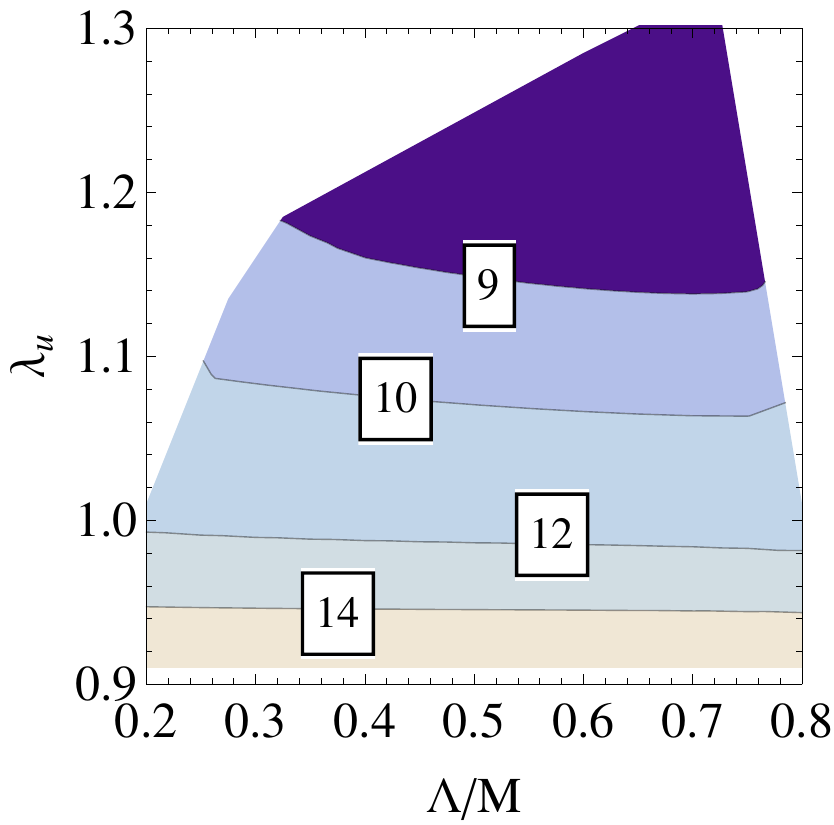}
\end{center}
\caption{Contour plots of $\log_{10}(M_{Landau} / {\rm GeV})$ for points with $m_h = 125$ GeV in typical MSSM models of ${\bf 5 \oplus \overline{5}}$ messengers with $N_{mess} = 4$ (left) and ${\bf 10 \oplus \overline{10}}$ messengers with $N_{mess} = 2$ (right).  The Landau pole is mainly caused by the blow up of  $\lambda_u$. }
\label{fig:landaupoles}
\end{figure}

The situation is somewhat different with respect to superpotential couplings. In ${\bf 5 \oplus \overline{5}}$ models $\lambda_u$ typically reaches a Landau pole before the GUT scale, since its value is necessarily quite large at the messenger scale and its RG evolution is dominated at one loop by $\lambda_u$ itself and also by $y_t$. In ${\bf 10 \oplus \overline{10}}$ models there are also large negative contributions from $g_3$, which help to control the running. These effects are evident in the $\beta$ functions, which in the MSSM case are dominated by
\begin{eqnarray}
\beta_{\lambda_u} &\sim& \frac{\lambda_{u}}{16 \pi^2} \left[ (N_{mess} + 3) \lambda_u^2 + 3 y_t^2 + \dots \right]  \hspace{2.2cm} \left( {\bf 5 \oplus \overline{5}} {\rm \; messengers} \right) \\
\beta_{\lambda_u} &\sim&   \frac{\lambda_{u}}{16 \pi^2} \left[ (3 N_{mess} + 3)\lambda_u^2 + 3 y_t^2 - \frac{16}{3} g_3^2 + \dots \right] \quad \quad \left( {\bf 10 \oplus \overline{10}} {\rm \; messengers} \right)  ~.
\end{eqnarray}
For  ${\bf 5 \oplus \overline{5}}$ messengers, there is always a Landau pole below the GUT scale in the range of parameters with $m_h = 125$ GeV. For ${\bf 10 \oplus \overline{10}}$ messengers there are Landau poles for $\lambda_u > 0.9$, while for $\lambda_u \leq 0.9$ the color contributions lead to an approximate fixed point. This is illustrated clearly in fig.\ \ref{fig:landaupoles}, which shows the scale of  the Landau pole in $\lambda_u$ across the parameter space with $m_h = 125$ GeV.  The occurrence of Landau poles is some sense a different manifestation of the same phenomenon that caused problems with radiative electroweak symmetry breaking: $\lambda_u$ must be large at the messenger scale to generate sizable $A$-terms. For  ${\bf 5 \oplus \overline{5}}$ messengers, the only two-loop contributions to $m_{H_u}^2$ and one-loop contributions to $\beta_{\lambda_u}$ are large and positive, while for ${\bf 10 \oplus \overline{10}}$ messengers there is a partial cancellation with color contributions in both the soft mass and beta function.

We emphasize, however, that the apparent Landau pole in $\lambda_u$ does not doom the models with ${\bf 5 \oplus \overline{5}}$ messengers. In a complete theory of dynamical supersymmetry breaking,  supersymmetry is broken by dimensional transmutation in a hidden sector gauge group. Perhaps the messengers at low scales are actually composites of the strong dynamics (as in theories of direct supersymmetry breaking), or are charged under the hidden sector gauge group and accumulate additional negative contributions to $\beta_{\lambda_u}$. In either case, the unknown strong dynamics  naturally control the apparent landau pole in $\lambda_u$, whose appearance is simply an artifact of maintaining the spurion limit all the way to the GUT scale. In this respect, fig.\ \ref{fig:landaupoles} indicates the scale at which new physics must appear in the hidden sector. Finally, although it is not entirely meaningful given the likely role of hidden sector dynamics, it is at least reassuring that there are models with ${\bf 10 \oplus \overline{10}}$ messengers and $m_h = 125$ GeV for which all couplings remain perturbative up to the GUT scale.

\bibliography{atermbib}
\bibliographystyle{JHEP}

\end{document}